\documentclass[a4paper,10pt]{article}
\usepackage[top=0.5in, bottom=1in, left=1in, right=1in]{geometry}
\usepackage{graphicx} 
\usepackage{natbib} 
\usepackage{color,amsmath,amssymb}
\usepackage{xcolor} 
\usepackage{float} 
\usepackage{geometry}
\usepackage{authblk}
\usepackage{soul}
\usepackage{algorithm}
\usepackage{algpseudocode}
\usepackage{hyperref}

\newcommand{\keywords}[1]{
    \noindent
    \textbf{Keywords:} #1
}

\begin{document}

\title{Estimating Media Mix Models with Demand–Marketing Interactions: A Constrained Genetic Algorithm Approach}

\author[a]{Jackie Siaw Tze Wong\footnote{Corresponding author}}
\author[c]{Gabriel Hughes}
\author[a]{Yanchun Bao}
\author[b]{Hongsheng Dai}
\author[a]{
Vasileios Giagos}
\author[c]{Fernanda Daniela Hinze Medina}
\author[c]{Hacer Oz Bakan}
\author[c]{Katherine Passmore}

\affil[a]{\small
School of Mathematics, Statistics, and Actuarial Science, University of Essex,  
Wivenhoe Park, Colchester, CO4 3SQ, UK. 
}

\affil[b]{School of Mathematics, Statistics and Physics, Herschel Building, Newcastle University, Newcastle upon Tyne, NE1 7RU, UK.}

\affil[c]{\small
Croud Inc Limited, The Bard Building, 20 Curtain Rd, London EC2A 3NG, UK.
}

\date{\today}

\maketitle

\begin{abstract}

    This paper proposes a novel extension to Media Mix Modeling (MMM) that introduces a multiplicative interaction between marketing activity and underlying consumer demand. Unlike standard MMM frameworks that assume additive and independent effects of media and baseline demand drivers, our specification allows marketing effectiveness to vary with prevailing demand conditions. However, the proposed structure introduces significant statistical challenges, particularly identifiability issues that can lead to unstable and biased parameter estimates. Using comprehensive simulation studies, we analyze the nature and severity of these identification issues through bias assessment and their implications for statistical inference. To address these issues, we develop a constrained genetic algorithm optimization approach which facilitates robust estimation that simultaneously mitigates biases arising from the aforementioned issues. The proposed approach also offers flexibility to incorporate economically meaningful parameter constraints. Finally, the proposed methodology is applied to real-world data to demonstrate its effectiveness in enhancing estimation accuracy while taking into account additional commercial constraints, facilitating budget allocation decisions.

\vspace{0.8cm}


\end{abstract}

\section{Introduction}
\label{sec:intro}

Media Mix Modeling (MMM) has undergone substantial development in recent years, driven by the increasing demand for privacy-preserving and data-driven approaches to marketing measurement. A seminal contribution is the Bayesian MMM framework of \cite{jin_bayesian_Carryover_Shape} from the Google Research Team, which formalized the incorporation of carryover and saturation effects through adstock transformations and nonlinear response functions, subsequently forming the basis of the LightweightMMM implementation and establishing many of the foundations of modern MMM. Building on these ideas, some other related developments from Google include \cite{sun2017geo,chen2018bias,zhang_bayesian_Reach_Frequency,zhang_bayesian_prior}. More recently, an industrial implementation, Google's Meridian framework, has evolved from the LightweightMMM framework by incorporating hierarchical modeling, causal inference extensions, and budget optimization capabilities (\cite{google2025meridian}). Meta's open-source Robyn framework (\cite{FacebookRobyn,runge2024packaging}) popularized the practical deployment of MMM through automated model selection, ridge regression, and evolutionary optimization techniques, making advanced MMM methodologies more accessible to practitioners. \cite{marin_bayesian_Shape_Funnel} proposed a Bayesian MMM framework incorporating physics-inspired response functions, including the Michaelis--Menten model, to provide a more interpretable representation of advertising saturation and consumer response dynamics. Other examples include \cite{sun2017geo,zhang_bayesian_prior,zhang_bayesian_Reach_Frequency}. Collectively, these developments reflect a broader trend towards increasingly sophisticated nonlinear MMM specifications that seek to better capture the complex mechanisms through which marketing activities influence business outcomes.

Despite substantial advances in MMM, an important limitation of many existing frameworks is their treatment of marketing activity and underlying consumer demand as additive and largely independent drivers of business outcomes. In practice, however, the effectiveness of marketing activity is often influenced by prevailing market conditions and consumer demand. The same level of advertising expenditure may generate markedly different outcomes depending on the level of underlying demand, suggesting the presence of interaction effects that are not adequately captured by conventional MMM specifications. This characterizes synergistic effects between consumer demand and media activities, which is closely related to the funnel effects discussed in \cite{chan_challenges_MMM}.

Introducing such interaction structures into MMM presents both opportunities and challenges. While multiplicative formulations can provide a more realistic representation of how marketing activities amplify or suppress existing demand, they also introduce additional model complexity. In particular, interaction-based specifications can suffer from identifiability issues, whereby multiple parameter combinations produce similar model fits. This can lead to unstable parameter estimates, biased attribution results, and difficulties in interpreting media effectiveness. Despite their practical importance, these estimation challenges have received relatively limited attention in the MMM literature.

At the same time, the estimation of increasingly complex MMM specifications requires optimization methods capable of handling nonlinear objective functions and parameter constraints. Traditional gradient-based estimation procedures may struggle in such settings, particularly when the likelihood surface exhibits multiple local optima or flat regions arising from identifiability concerns. Evolutionary optimization techniques, such as genetic algorithms (GAs), offer a flexible alternative by enabling global search over constrained parameter spaces without relying on analytical derivatives. Evolutionary optimization methods are increasingly used in MMM. Meta’s Robyn framework (\citealp{runge2024packaging}) applies multi-objective evolutionary optimization via Nevergrad for hyperparameter calibration, while \cite{dulara2025machine} employ GAs for constrained media allocation. However, these approaches focus on hyperparameter tuning or post-estimation optimization rather than direct parameter estimation. To the best of our knowledge, GAs have not been used for explicit estimation of the MMM parameters.

Motivated by these challenges, this paper makes four primary contributions. First, we propose a novel MMM framework that incorporates a multiplicative interaction between media variables and underlying demand, allowing advertising effectiveness to vary according to prevailing market conditions. Second, through extensive simulation studies, we demonstrate the identifiability challenges associated with this model specification and quantify their impact on parameter estimation and media attribution. Third, we propose a GA-based constrained estimation procedure that not only serves as an efficient optimization tool, but also enables the incorporation of economically meaningful constraints that help mitigate identifiability issues inherent in the proposed model. Finally, we illustrate the practical applicability of the proposed methodology using a real-world case study. 

The rest of the paper is organized as follows. The remainder of Section \ref{sec:intro} presents a brief overview of the traditional MMM framework. Section \ref{sec:model} develops the proposed variant of MMM specification and highlights the differences with the traditional framework. Section \ref{sec:sim_study} investigates the statistical behavior of conventional estimates through simulation and highlights the potential biases arising from identifiability issues. Section \ref{sec:genetic} outlines the proposed genetic optimization algorithm tailored to suit our implementations and evaluates its performance relative to conventional optimization approaches. Section \ref{sec:econ} extends the framework to incorporate commercially relevant constraints. Empirical results are reported in Section \ref{sec:empirical}. Finally, we conclude with some closing remarks, discuss limitations and avenues for future research in Section \ref{sec:conclusion}.


\subsection{Traditional MMM}

The traditional MMM framework (see for example \cite{jin_bayesian_Carryover_Shape}) specifies an additive regression model:
\begin{equation}
Y_t = \tau + \boldsymbol X_t^\top \boldsymbol \beta_X + \boldsymbol Z_t^\top \boldsymbol \beta_Z + \epsilon_t\ ,
\label{eq:traditional_MMM}
\end{equation}
where $Y_t$ denotes the response variable, namely the Key Performance Indicator (KPI), at time $t$ (in weeks\footnote{Time may be measured in days, weeks, or months; we use weeks in our study to reflect common practice.}), $\tau$ is the (non-temporal) baseline sales, $\boldsymbol \beta_X$ and $\boldsymbol \beta_Z$ are the regression coefficients, $\boldsymbol X_t$ denotes the media variables, $\boldsymbol Z_t$ denotes the non-media variables (also known as the control variables), and $\epsilon_t$ is white noise which has constant variance. In essence, the traditional MMM framework specifies a linear additive relationships between the response and baseline, media, control variables.

\subsection{The media variable}
\label{sec:media_variable}

Each component of the media variable $\boldsymbol X_t = (X_{1,t}, \ldots, X_{M,t})$ represents the effectiveness of a corresponding media channel $m \in \{1, \ldots, M\}$ at time $t$. The total media contribution at time $t$ is therefore $\boldsymbol X_t^\top \boldsymbol \beta_X = \sum_{m=1}^{M} X_{m,t} \beta_{X,m}$. Modeling $X_{m,t}$ typically involves two sequential nonlinear transformations: first, accounting for the carryover effects (adstock) and second, saturation effects (diminishing returns).

\subsubsection{Carryover effects}

The impact of advertising often extends beyond the immediate exposure period, a phenomenon known as the carryover effects. We model this using the adstock transformation, which is essentially a first-order temporal decay function where the current adstocked impressions $A_{m,t}$ for media channel $m$ depend on both the current media activity and the previous adstocked impressions. Let $X_{m,t}^{\text{raw}}$ denote the raw media activity (for instance, impressions or spending) for channel $m$ at time $t$. The adstocked impressions $A_{m,t}$ for $t=1,\ldots, T$ are computed as:
\begin{equation}
\begin{array}{l l}
A_{m,t} = X_{m,t}^{\text{raw}} + \alpha_m \cdot A_{m,t-1} \ ,
\end{array}
\label{eq:adstock}
\end{equation}
where $A_{m,0}=0$, $A_{m,t-1}$ denotes the adstocked impressions from the previous period, and $\alpha_m \in (0, 1)$ is the adstock rate for channel $m$ which represents the decay rate of marketing effectiveness from one period to the next (a higher value indicates a longer-lasting carryover effect). This formulation is a simplified geometric decay model, in which the influence of past media activity diminishes exponentially over time.

\subsubsection{Saturation effects}

Following the adstock transformation, the adstocked media activity $A_{m,t}$ is then subjected to a nonlinear shape transformation function to capture the phenomenon of diminishing returns or saturation effects (effectiveness of additional advertising spend eventually plateaus). Following \cite{jin_bayesian_Carryover_Shape}, we employ the \textbf{Hill function} for this purpose:

\begin{equation}
X_{m,t} = \text{Hill}(A_{m,t}, \gamma_m, \kappa_m) = \frac{A_{m,t}^{\gamma_m}}{A_{m,t}^{\gamma_m} + \kappa_m^{\gamma_m}}.
\label{eq:hill_function}
\end{equation}
Here, $X_{m,t}$ is the final transformed effectiveness variable for channel $m$ at time $t$ that enters the main model \eqref{eq:traditional_MMM}. The parameters are:
\begin{itemize}
    \item $\gamma_m \in (0, \infty)$: The \textbf{shape parameter}, which determines the overall curvature of the response:
    \begin{itemize}
        \item If $\gamma_m \in (0, 1]$, the function increases monotonically and exhibits a concave (L-shaped) curve, implying diminishing returns from the outset.
        \item If $\gamma_m \in (1, \infty)$, the function displays an S-shaped curve (convex initially, then concave), suggesting an initial “warming-up” phase where effectiveness increases, followed by diminishing returns. 
    \end{itemize}
    \item $\kappa_m \in (0, \infty)$: The \textbf{inflection parameter}, which represents the point at which the growth rate of the response begins to slow and diminishing returns become more pronounced. This is often interpreted as the media activity level required to achieve half of the maximum potential effectiveness. A larger $\kappa_m$ indicates that a higher level of adstocked activity is needed to reach this saturation point.
\end{itemize}

\begin{figure}[htbp]
\centering
\includegraphics[scale=0.5]{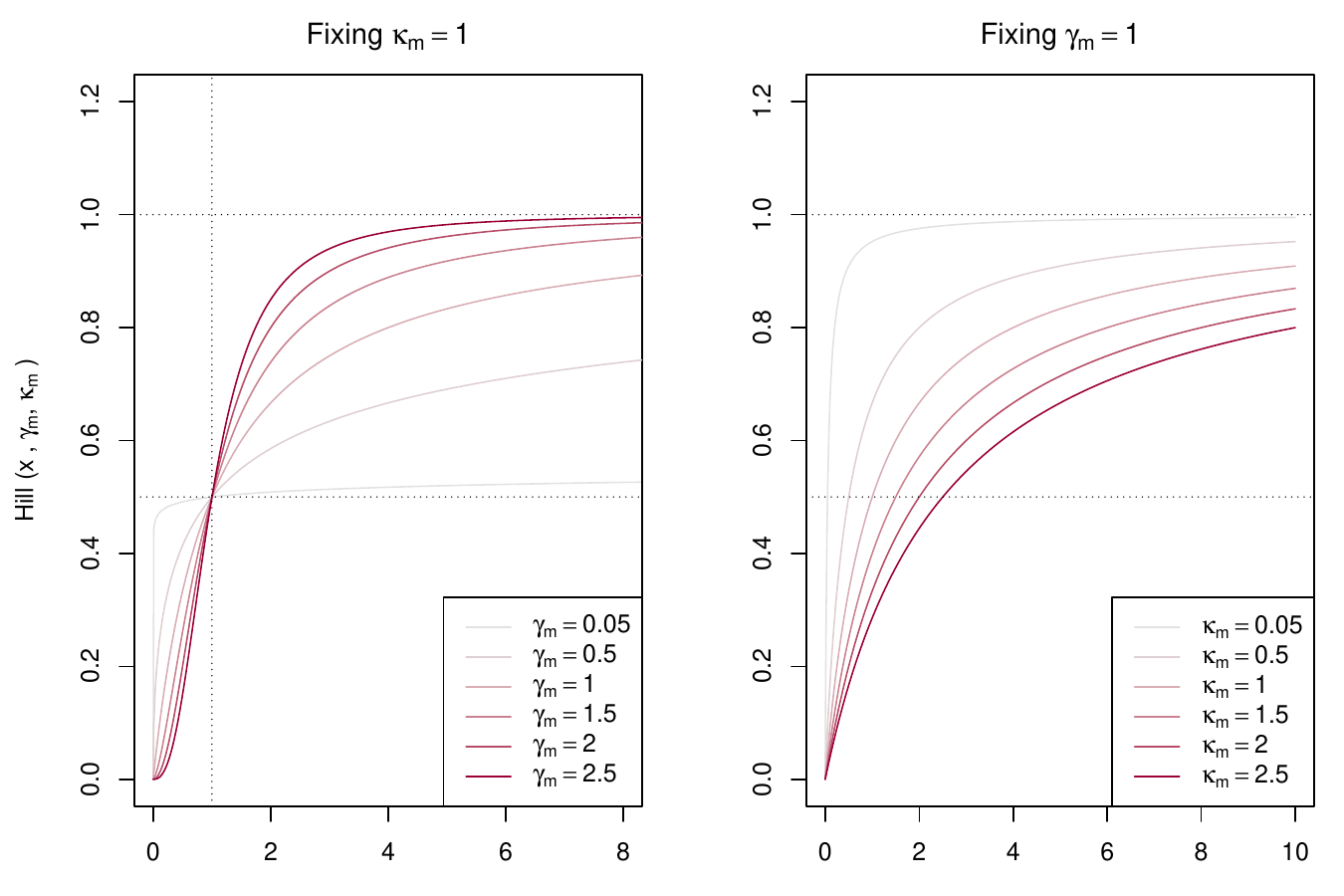}
\caption{\label{fig:hill_fn_k_and_s}The Hill function plotted as a function of the shape parameter $\gamma_m$ and the inflection parameter $\kappa_m$. Left: $\gamma_m$ varies with $\kappa_m = 1$; Right: $\kappa_m$ varies with $\gamma_m = 1$.}
\end{figure}

\begin{figure}[htbp]
\centering
\includegraphics[scale=0.5]{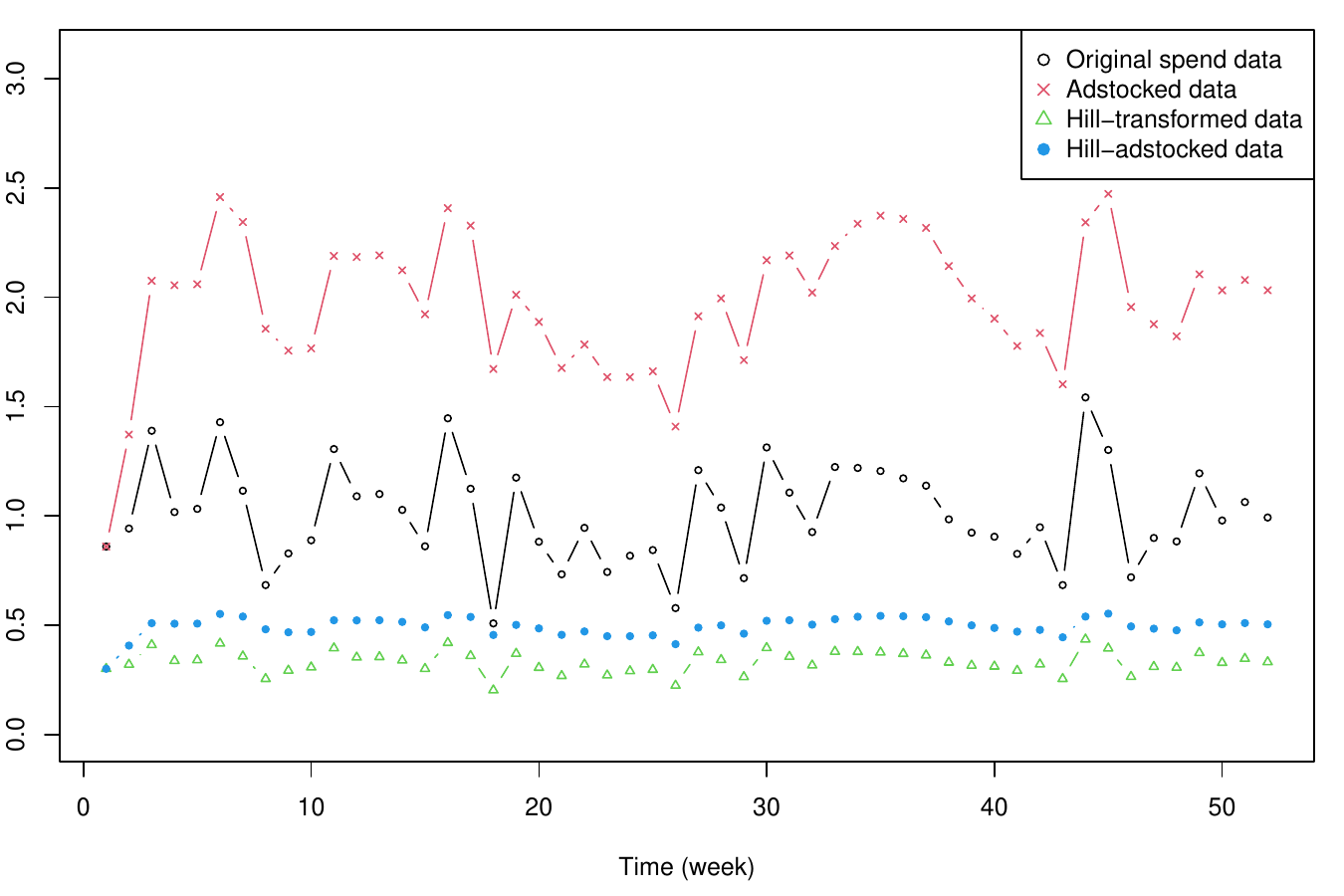}
\caption{\label{fig:hill_adstock_sim}Simulated weekly media spend data and the corresponding transformed data using the adstock function in (\ref{eq:adstock}), Hill function in (\ref{eq:hill_function}), and both, where we set $\alpha_m=0.5$, $\gamma_m=1$, $\kappa_m=2$. The weekly raw data $X^{\text{raw}}_{m,t}$ ($t=1,\ldots,52$) is simulated from $N(1,0.25^2)$.}
\end{figure}

The Hill function, bounded between 0 and 1 (before being scaled by the $\boldsymbol \beta_X$ coefficients in the main model), provides flexibility to model various response patterns. The left panel of Figure \ref{fig:hill_fn_k_and_s} illustrates the effect of varying $\gamma_m$ while keeping $\kappa_m$ constant, showing how the shape changes from L-shaped to S-shaped. The right panel demonstrates the impact of $\kappa_m$ on shifting the saturation point, with $\gamma_m$ held fixed. Figure \ref{fig:hill_adstock_sim} presents simulated examples of how raw media activity, $X_{m,t}^{\text{raw}}$, is transformed by both the adstock and Hill functions to form $X_{m,t}$.


\section{Model specification}
\label{sec:model}

\subsection{An alternative MMM specification}

In this paper, we propose the following MMM framework:
\begin{equation}
Y_t = \left(\boldsymbol U_t^\top \boldsymbol \beta_U\right) \cdot \left(1 + \boldsymbol X_t^\top \boldsymbol \beta_X\right) + \boldsymbol Z_t^\top \boldsymbol \beta_Z + \epsilon_t
\label{eq:model}
\end{equation}
for $t=1,\ldots,T$. The residual $\epsilon_t$ is assumed to follow an unknown distribution $F$ with mean $0$ and variance $\sigma^2$. At its core, this is a basic regression model, where the term $\boldsymbol U_t^\top \boldsymbol \beta_U \cdot \boldsymbol X_t^\top \boldsymbol \beta_X$ is analogous to interaction effects in standard statistical modeling, that the impact of one set of covariates depends on another.


In context of Media Mix Modeling, the response variable is hypothesized to depend on a set of predictors comprising three main components: a vector variable $\boldsymbol U_t$ representing the underlying demand (baseline sales), a vector variable $\boldsymbol X_t$ capturing the effects of various media channels, and a vector $\boldsymbol Z_t$ representing the effects non-media or control variables. The term $\boldsymbol U_t^\top \boldsymbol \beta_U$ captures the effects of the underlying demand. A key feature of our MMM specification is the multiplicative interaction between underlying demand and various media effects, $\boldsymbol U_t^\top \boldsymbol \beta_U \cdot \boldsymbol X_t^\top \boldsymbol \beta_X$, which represents the incremental lift attributable to media activities. This formulation implies that media effectiveness scales with the magnitude of baseline demand, reflecting a context-dependent (rather than static) impact. For instance, during periods of elevated underlying demand, such as Black Friday, the incremental impact of media on the KPI is amplified. This aligns with the empirical observation that marketing efforts often yield higher returns when baseline consumer interest or purchasing intent is already strong.

Note that the function $\boldsymbol X_t = \boldsymbol X_t(\boldsymbol \alpha, \boldsymbol \gamma, \boldsymbol \kappa,\boldsymbol{X}_1^{\text{raw}},\ldots,\boldsymbol{X}_t^{\text{raw}})$ explicitly indicates that the transformed media effectiveness variables $\boldsymbol X_t$ are dependent on the adstock rates, shape parameters, and inflection parameters, as defined in \eqref{eq:adstock} and \eqref{eq:hill_function}, and all previous raw media variables. For brevity, we suppress the dependence and write only $\boldsymbol X_t$.

The non-media (control) variable $\boldsymbol Z_t$ usually represents factors induced by major external shocks, such as the COVID-19 pandemic. These effects can be proxied using indicators such as the Google Mobility Index (specifically its residential and retail components) or policy-related variables such as lockdown measures. 

Our proposed MMM framework differs from the traditional MMM specification in (\ref{eq:traditional_MMM}) in three aspects:
\begin{itemize}
    \item the baseline sales are assumed to be time-varying rather than being static across weeks;
    \item the baseline sales are associated with regression coefficients to calibrate their effects with the response;
    \item the multiplicative relationship between the baseline sales and media variables.
\end{itemize}

\subsection{Conventional estimations}

Suppose we denote the parameter vector as $\boldsymbol \theta \in \mathbb{R}^p$ ($p=p_U+p_X+p_Z$ is the total number of model parameters), where $p_U$ and $p_Z$ are respectively the total number of variables in $\boldsymbol{U}_t$ and $\boldsymbol{Z}_t$, and $p_X=4\times M$ ($\beta_{X,m}$, $\alpha_m$, $\gamma_m$, $\kappa_m$ for $m=1,\ldots,M$) is the total number of variables in $\boldsymbol X_t$.

Here, the parameter vector $\boldsymbol \theta$ encompasses all individual coefficients and nonlinear transformation parameters:
\begin{equation}
\boldsymbol \theta = \left( \boldsymbol \beta_U^\top, \boldsymbol \beta_X^\top, \boldsymbol \beta_Z^\top, \boldsymbol \alpha^\top, \boldsymbol \gamma^\top, \boldsymbol \kappa^\top \right)^\top
\end{equation}
where:
\begin{itemize}
    \item $\boldsymbol \beta_U$ represents the coefficients for the underlying demand components;
    \item $\boldsymbol \beta_X$ represents the coefficients for the media effectiveness;
    \item $\boldsymbol \beta_Z$ represents the coefficients for the control variables;
    \item $\boldsymbol \alpha = (\alpha_1, \ldots, \alpha_M)^\top$ is the vector of adstock rates for each of the $M$ media channels;
    \item $\boldsymbol \gamma = (\gamma_1, \ldots, \gamma_M)^\top$ is the vector of shape parameters for each media channel's Hill function;
    \item $\boldsymbol \kappa = (\kappa_1, \ldots, \kappa_M)^\top$ is the vector of inflection parameters for each media channel's Hill function.
\end{itemize}

In conventional estimations, the objective is to find $\boldsymbol \theta$ that minimizes the discrepancy between the observed KPI ($Y_t$) and the model's predictions, in the form of either least squares or maximum likelihoods.

\subsubsection{Least Squares Estimation (LSE)}

The LSE approach aims to minimize the squared model residuals to find the optimal parameters. Mathematically, 
\begin{equation}
\widehat {\boldsymbol \theta} = \arg\min_{\boldsymbol \theta} \sum_{t=1}^T \left(Y_t - \left[ \left(\boldsymbol U_t^\top \boldsymbol \beta_U\right) \cdot \left(1 + \boldsymbol X_t^\top \boldsymbol \beta_X\right) + \boldsymbol Z_t^\top \boldsymbol \beta_Z \right]\right)^2.
\label{eqn:LSE}
\end{equation}
The variance parameter is then estimated as 
\begin{equation}
    \widehat{\sigma}^2 = \frac{1}{T-p}\sum_{t=1}^T \left(Y_t - \left[ \left(\boldsymbol U_t^\top \boldsymbol{\widehat{\beta}}_U\right) \cdot \left(1 + \boldsymbol X_t^\top \boldsymbol{\widehat{\beta}}_X\right) + \boldsymbol Z_t^\top \boldsymbol{\widehat{\beta}}_Z \right]\right)^2.
\end{equation}

\subsubsection{Maximum Likelihood Estimation (MLE)}

The MLE approach aims to maximize the log-likelihood function to find the optimal parameters. Mathematically, 
\begin{equation}
\{\widehat {\boldsymbol \theta},\widehat{\sigma}^2\} =\arg\max_{\boldsymbol \theta,\sigma^2} \sum_{t=1}^T \left\{ -\frac{1}{2}\log(2\pi \sigma^2) -\frac{1}{2\sigma^2}\left(Y_t - \left[ \left(\boldsymbol U_t^\top \boldsymbol \beta_U\right) \cdot \left(1 + \boldsymbol X_t^\top \boldsymbol \beta_X\right) + \boldsymbol Z_t^\top \boldsymbol \beta_Z \right]\right)^2 \right\}.
\label{eqn:MLE}
\end{equation}

\subsection{Identifiability issues}
\label{sec:identifiability}

The nonlinear transformations within the MMM framework present significant challenges in the model estimation stage, due to the weak identifiability of media-related parameters, as noted by \cite{jin_bayesian_Carryover_Shape}. In particular, the authors demonstrated that various combinations of $\boldsymbol{\alpha}$, $\boldsymbol{\gamma}$, and $\boldsymbol{\kappa}$ can produce very similar media response functions over the observed range of advertising expenditure. Consequently, these parameters may be only weakly identified in practice, with the likelihood surface exhibiting relatively flat regions that make it difficult to distinguish between competing parameter values. In what follows, we will numerically demonstrate the biases induced by such identifiability issues through simulation studies. 

\section{Simulation studies}
\label{sec:sim_study}

In this section, we present simulation results to illustrate the performance of conventional estimation approaches in our setting. As will become evident, the nonlinear transformations in the MMM give rise to identifiability issues, causing conventional estimation approaches to fail to correctly and reliably recover the true model parameters.

\subsection{Simulation configurations}

Data of size $T=1,000$ are generated using the following simulation settings for $t=1,\ldots,T$:
\begin{itemize}
    \item $U_{i,t}\sim N(0,1)$ for $i=1,\ldots,6$; 
    \item $\log X^{\text{raw}}_{m,t}\sim N(0,1)$ for $m=1,\ldots,M=3$;
    \item $Z_{i,t}\sim N(0,1)$ for $i=1,\ldots,4$;
\end{itemize}
so $p_U=6$, $p_X=4\times M=12$, $p_Z=4$, and thus, $p=22$.

We consider the following two configurations of the true parameter values. 
\begin{itemize}
    \item M1: MMM without nonlinear transformations.
     \begin{eqnarray}
\left\{
\begin{array}{l}
\boldsymbol{\beta}_U=(-1,1,0.5,2,5,-3)^\top, \\
\boldsymbol{\beta}_X=(-0.5,1,2)^\top, \\
\boldsymbol{\beta}_Z=(-1,0.5,1,6)^\top, \\
\sigma = 5, \\
\boldsymbol{X}_t = \boldsymbol{X}^{\text{raw}}_t.
\end{array}
\right.
\label{eqn:M1_true_param}
     \end{eqnarray}
    \item M2: MMM with nonlinear transformations.
         \begin{eqnarray}
\left\{
\begin{array}{l}
\boldsymbol{\beta}_U=(-1,1,0.5,2,5,-3)^\top, \\
\boldsymbol{\beta}_X=(-0.5,1,2)^\top, \\
\boldsymbol{\beta}_Z=(-1,0.5,1,6)^\top, \\
\boldsymbol{\alpha}=(0.5,0.5,0.5)^\top, \\
\boldsymbol{\gamma}=(3,3,3)^\top, \\
\boldsymbol{\kappa}=(1,1,1)^\top, \\
\sigma = 5, \\
\boldsymbol{X}_t \mbox{ is obtained from } \boldsymbol{X}^{\text{raw}}_t \mbox{ using (\ref{eq:adstock}) and (\ref{eq:hill_function})}.
\end{array}
\right.
\label{eqn:M2_true_param}
     \end{eqnarray}
\end{itemize}

\subsection{Simulation results}
\label{sec:sim_results}

We first conduct the experiment for a single simulation run and examine the estimation accuracy of both the MLE and LSE approaches in detail. The corresponding results are reported in Tables \ref{tab:M1_tab_result} and \ref{tab:M2_tab_result}, where the mean absolute differences (MAD),
\begin{equation}
    \operatorname{MAD} = \frac{1}{p}\times\sum_{j=1}^{p}\left|\theta_j-\widehat{\theta}_j\right|, 
\end{equation}
are also included to quantify overall differences from the true values.

In the M1 configuration, both MLE and LSE accurately recover the model parameters, with deviations attributable to the usual finite-sample simulation variability. In contrast, for the configuration M2, the estimated parameters exhibit substantial bias in most cases, with the exception of a few parameters such as $\boldsymbol{\widehat{\beta}}_Z$ and $\widehat{\sigma}$. The most notable discrepancies arise in the media-specific parameters, which are estimated particularly poorly and display large bias magnitudes. For instance, $\beta_{X,1}$ is severely underestimated, whereas $\beta_{X,2}$ is considerably overestimated. These results provide preliminary evidence of the identifiability challenges associated with M2.

\begin{table}[ht]
\centering
\small
\caption{Estimated parameters of M1 using MLE and LSE approaches.}
\label{tab:M1_tab_result}
\begin{tabular}{rrrr}
  \hline
 & TRUE & MLE & LSE \\ 
  \hline
  $\beta_{U,1}$ & -1.00000 & -1.02573 & -1.02553 \\ 
  $\beta_{U,2}$ & 1.00000 & 1.05726 & 1.05702 \\ 
  $\beta_{U,3}$ & 0.50000 & 0.47019 & 0.47006 \\ 
  $\beta_{U,4}$ & 2.00000 & 2.03014 & 2.02968 \\ 
  $\beta_{U,5}$ & 5.00000 & 5.09370 & 5.09258 \\ 
  $\beta_{U,6}$ & -3.00000 & -3.05363 & -3.05296 \\ 
  $\beta_{X,1}$ & -0.50000 & -0.50565 & -0.50572 \\ 
  $\beta_{X,2}$ & 1.00000 & 1.00010 & 1.00036 \\ 
  $\beta_{X,3}$ & 2.00000 & 1.95480 & 1.95526 \\ 
  $\beta_{Z,1}$ & -1.00000 & -1.27945 & -1.27927 \\ 
  $\beta_{Z,2}$ & 0.50000 & 0.36193 & 0.36176 \\ 
  $\beta_{Z,3}$ & 1.00000 & 0.98682 & 0.98743 \\ 
  $\beta_{Z,4}$ & 6.00000 & 5.89219 & 5.89156 \\ 
  $\sigma$ & 5.00000 & 4.973332 & 5.006064 \\ 
  \hline
  MAD &  & 0.06474 & 0.06308 \\
   \hline
\end{tabular}
\end{table}

\begin{table}[ht]
\centering
\small
\caption{Estimated parameters of M2 using MLE and LSE approaches.}
\label{tab:M2_tab_result}
\begin{tabular}{rrrr}
  \hline
 & TRUE & MLE & LSE \\ 
  \hline
  $\beta_{U,1}$ & -1.00000 & -1.67246 & -1.59125 \\ 
  $\beta_{U,2}$ & 1.00000 & 1.82877 & 1.74061 \\ 
  $\beta_{U,3}$ & 0.50000 & 0.67560 & 0.64385 \\ 
  $\beta_{U,4}$ & 2.00000 & 3.48460 & 3.31818 \\ 
  $\beta_{U,5}$ & 5.00000 & 8.80858 & 8.38258 \\ 
  $\beta_{U,6}$ & -3.00000 & -5.28647 & -5.03079 \\ 
  $\beta_{X,1}$ & -0.50000 & -8.83319 & -7.70470 \\ 
  $\beta_{X,2}$ & 1.00000 & 9.15313 & 8.11766 \\ 
  $\beta_{X,3}$ & 2.00000 & 0.73071 & 0.76463 \\ 
  $\alpha_1$ & 0.50000 & 0.343933 & 0.375217 \\ 
  $\alpha_2$ & 0.50000 & 0.529501 & 0.528544 \\ 
  $\alpha_3$ & 0.50000 & 0.472088 & 0.471395 \\ 
  $\gamma_1$ & 3.00000 & 0.714352 & 0.948086 \\ 
  $\gamma_2$ & 3.00000 & 0.958592 & 0.954822 \\ 
  $\gamma_3$ & 3.00000 & 5.114809 & 5.152541 \\ 
  $\kappa_1$ & 1.00000 & 0.002363 & 0.011043 \\ 
  $\kappa_2$ & 1.00000 & 0.044866 & 0.053470 \\ 
  $\kappa_3$ & 1.00000 & 1.217172 & 1.21867 \\ 
  $\beta_{Z,1}$ & -1.00000 & -1.29190 & -1.29142 \\ 
  $\beta_{Z,2}$ & 0.50000 & 0.38810 & 0.38874 \\ 
  $\beta_{Z,3}$ & 1.00000 & 1.00121 & 1.00061 \\ 
  $\beta_{Z,4}$ & 6.00000 & 5.87288 & 5.87374 \\ 
  $\sigma$ & 5.00000 & 4.948082 & 5.002961 \\ 
   \hline
    MAD &  & 1.59468 & 1.44261 \\
    \hline
\end{tabular}
\end{table}

\begin{figure}[htbp]
\centering
\includegraphics[scale=0.35]{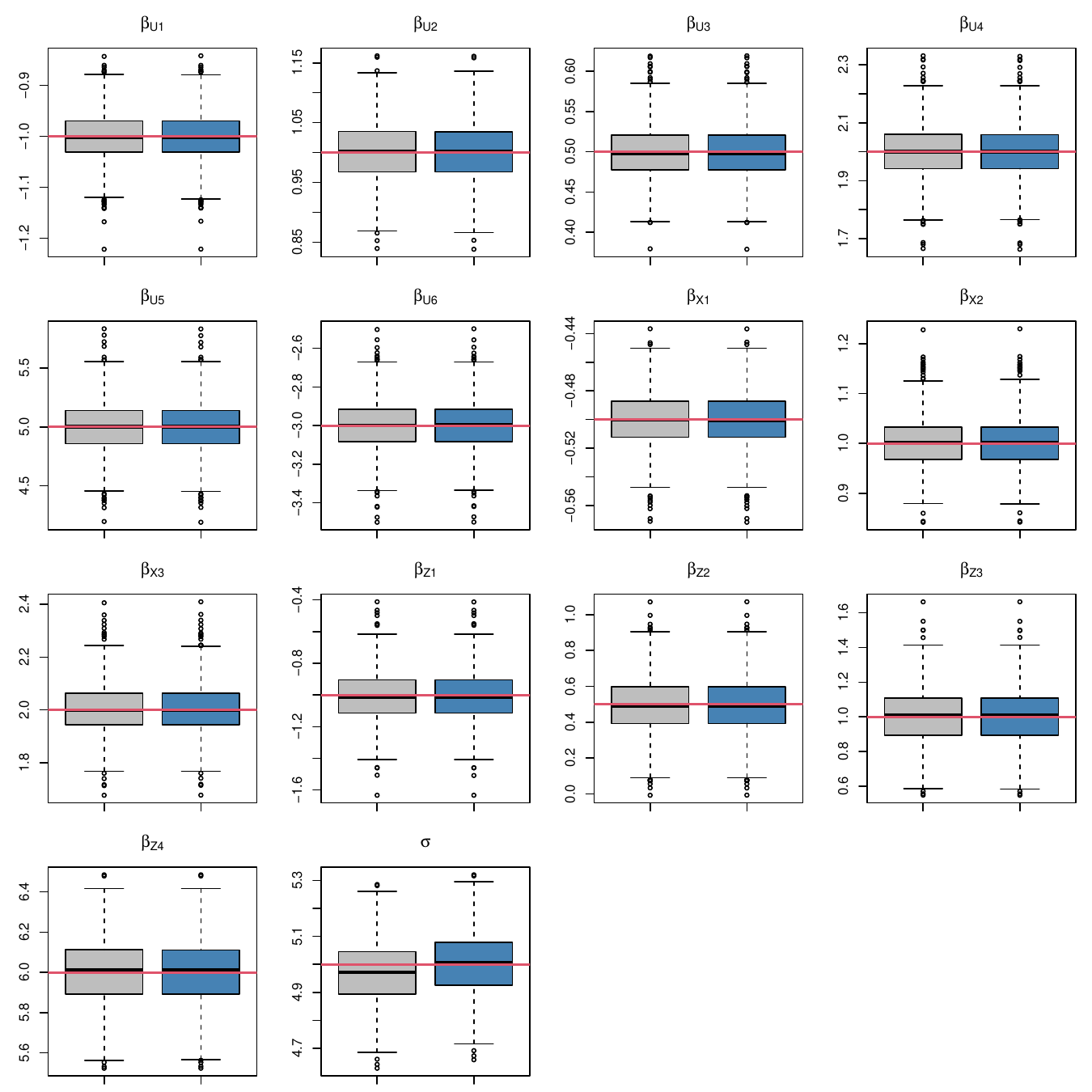}
\caption{\label{fig:M1_boxplot_result}Boxplots of the estimated parameters under M1. True parameter values are indicated by red horizontal lines. Grey and blue boxes correspond to the MLE and LSE estimates, respectively.}
\end{figure}

\begin{figure}[htbp]
\centering
\includegraphics[scale=0.375]{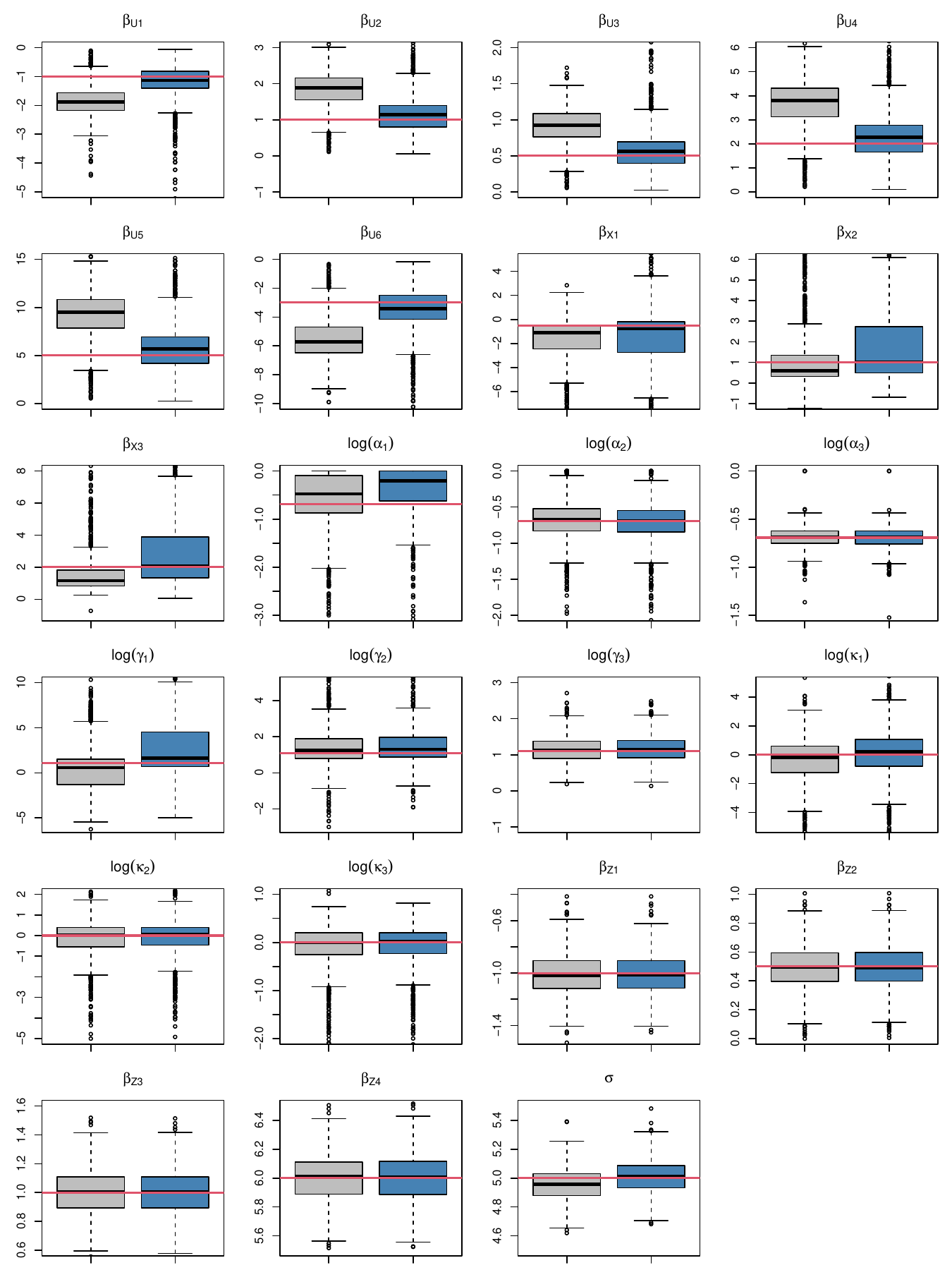}
\caption{\label{fig:M2_boxplot_result}Boxplots of the estimated parameters under M2. True parameter values are indicated by red horizontal lines. Grey and blue boxes correspond to the MLE and LSE estimates, respectively.}
\end{figure}

The experiments are then repeated $R=1,000$ times to derive the statistical properties of the estimates. Figures \ref{fig:M1_boxplot_result} and \ref{fig:M2_boxplot_result} illustrate the boxplots of estimated parameters under the M1 and M2 configurations respectively. Evidently, the estimates under the M1 configuration are again unbiased, where both LSE and MLE approaches correctly identify all of the model parameters, including the variance parameter. By contrast, the estimates obtained for M2 are broadly biased. In particular, the estimates of a number of the model parameters are not only biased (e.g. $\boldsymbol{\widehat{\beta}}_{U}$, $\boldsymbol{\widehat{\beta}}_{X}$, $\alpha_1$), but also contain large deviations with many outliers, indicating that estimates produced are mostly unstable and unreliable.

The non-robustness of the estimates observed here is a well-documented issue in MMM frameworks, as discussed in Section \ref{sec:identifiability}. These estimation biases stem from weakly identified parameters induced by nonlinear transformations, which manifest as flat or poorly behaved objective functions in the estimation procedure. Although our model specification differs slightly from the standard MMM formulation, the same identification issues persist, as evidenced in our simulation study. 

\section{Genetic Algorithm (GA)}
\label{sec:genetic}

The basic idea behind the GA is to generate candidate solutions (population members) through stochastic evolutionary operators such as breeding and mutation, with the aim of identifying candidates with superior characteristics as measured by the objective score. In essence, the key to having high statistical efficiency is to have high genetic diversity to ensure that the search is wide enough to span the parameter space, where relevant. Through generations of ``evolution'', the population is eventually at a saturation point, where all members are elites with good characteristics. For more details, see \cite{whitley1994genetic}.

In this spirit, a summary of our proposed GA is:
\begin{itemize}
\item Start with an initial population.
\item Members of the initial population are allowed to breed and mutate, producing offspring and mutated members. 
\item The population members are evaluated using a certain scoring system. A portion of the population is retained as ``elites'', the rest is eliminated.
\item The elites are then allowed to breed and mutate, producing new offspring and newly mutated members.
\item They are evaluated using the same scoring system, giving a new set of elites.
\item Then repeat for many iterations.
\item Assuming the algorithm is constructed and tuned appropriately, convergence will be attained.
\end{itemize}

\subsection{Initialization}
\label{sec:init}

The initial population is generated from an initial parameter vector $\boldsymbol{\theta}^{(0)}$. Specifically, each individual in the population is constructed as a perturbed copy of 
$\boldsymbol{\theta}^{(0)}$, where random perturbations are introduced to promote diversity in the search process. Two types of perturbation are used:

\begin{enumerate}
    
\item Multiplicative perturbation for level-type parameters (i.e.\ the regression coefficients $\beta_{\cdot,\cdot}$). For a component of the parameter $\theta_j$:
\[
\theta_j \leftarrow \theta_j \times U(a,b),
\]
where \(U(a,b)\) denotes a uniform random variable.

\item Additive perturbation for log-scale parameters (i.e.\ the media-specific parameters $\alpha_m,\gamma_m,\kappa_m$ $(m=1,\ldots,M)$ and $\sigma$):
\[
\theta_j \leftarrow \theta_j + \varepsilon_{j,\text{init}},
\qquad
\varepsilon_{j,\text{init}} \sim N(0,\sigma_{\text{init}}^2).
\]
\end{enumerate}
The perturbation scales can be chosen differently across parameter groups to allow both local exploration and wider coverage of the parameter space.

Repeating this procedure produces an initial population set,
\[
\mathbf{P}^{(0)}
=
\{\boldsymbol{\theta}^{(1)},\ldots,\boldsymbol{\theta}^{(n_0)}\},
\]
where \(n_0\) is the initial population size. The first individual is kept equal to the original parameter vector \(\boldsymbol{\theta}^{(1)}=\boldsymbol{\theta}^{(0)}\) so that the baseline solution is always retained.

Clearly, the initial population is important to control initial genetic diversity, which is one of the governing factors for how wide the search space is for the parameter.

\subsection{Mechanisms for population dynamics}
\label{sec:pop_dynamics}

Once the initial population has been established, we will simulate how the population emerges over time using certain population dynamics. Here, we define mainly two mechanisms, breeding/reproduction and mutation. The corresponding rates of breeding and mutation are denoted by $r_b$ and $r_m$, respectively.

\subsubsection{Breeding/reproduction and crossover}
\label{sec:breeding}

New offspring are generated from two parent parameter vectors,
\(\boldsymbol{\theta}^{(1)}\) and \(\boldsymbol{\theta}^{(2)}\). For each parameter component \(\theta_j\), one of the following crossover mechanisms is randomly selected.

\begin{itemize}
\item Direct Inheritance. The offspring may inherit the parameter directly from one of the parents:
\begin{equation}
   \theta_j^{(\text{child})} = \left\{\begin{array}{ll}
     \theta_j^{(1)}, & \mbox{ with probability 0.25,} \\
     \theta_j^{(2)}, & \mbox{ with probability 0.25.} \\
    \end{array}\right.
\end{equation}

\item Arithmetic Crossover. With probability 0.25, a weighted average of the parental parameters is generated:
\[
\theta_j^{(\text{child})}
=
\pi \theta_j^{(1)}
+
(1-\pi)\theta_j^{(2)},
\]
where
\[
\pi \sim U(0,1).
\]
This allows interpolation between the two parent chromosomes.

\item BLX-\(\alpha\) Crossover (see \cite{herrera1998tackling}). To encourage broader exploration, a blended crossover (BLX-\(\alpha\)) operator is used. Let the component-wise absolute difference be
\[
d_j = \left|\theta_j^{(1)} - \theta_j^{(2)}\right|.
\]
With probability 0.25, the offspring parameter is sampled from
\[
\theta_j^{(\text{child})}
\sim
U\left(
\min\left\{\theta_j^{(1)},\theta_j^{(2)}\right\} - \delta d_j,\,
\max\left\{\theta_j^{(1)},\theta_j^{(2)}\right\} + \delta d_j
\right),
\]
where \(\delta > 0\) controls the exploration range.
\end{itemize}

Finally, small additive and multiplicative random perturbations (similar to Section \ref{sec:init}) are added to the offspring parameters to maintain population diversity and avoid premature convergence.

\subsubsection{Mutation}
\label{sec:mutation}

Mutation is applied to each component of the parameter vector 
\(\boldsymbol{\theta}\) to maintain genetic diversity throughout the optimization process. For each parameter component \(\theta_j\), one of the following mutation operations is randomly selected.

\begin{itemize}
\item Gaussian mutation. With probability \(0.50\), a Gaussian perturbation is applied:
\[
\theta_j \leftarrow \theta_j + \varepsilon_{j,\text{mutate}},
\qquad
\varepsilon_j \sim N(0,\sigma_{j,\text{mutate}}^2),
\]
where \(\sigma_{j,\text{mutate}}\) is a parameter-specific mutation scale which should be carefully tuned to promote diversity. 

\item No mutation. With probability \(0.45\), the parameter remains unchanged:
\[
\theta_j \leftarrow \theta_j.
\]

\item Sign-flip mutation. With a small probability \(0.05\), a sign-flip mutation is applied:
\[
\theta_j \leftarrow -\theta_j.
\]
This occasional large perturbation helps the algorithm escape local optima and improves exploration of the parameter space.
\end{itemize}

\subsection{Synthesizing population}

Combining the simulation processes described in Sections \ref{sec:init} and \ref{sec:pop_dynamics} provides a procedure for synthesizing a population from a given initial parameter vector $\boldsymbol{\theta}^{(0)}$ for a number of generations, $g$, through breeding/crossover and mutation, giving a population of size
\[
n_0(r_b+r_m)^g.
\] In our study, we simulate the population through $g=5$ generations of breeding and mutation in each iteration.


\subsection{Evaluation metrics}
\label{sec:genetic_eval}

As inspired by \cite{konak2006multi}, we propose a multi-objective score function, $S(\widehat{\boldsymbol{\theta}})$, which is a weighted average of several key statistical and constraint satisfaction metrics. This approach allows for a flexible prioritization of different model quality aspects. Specifically, for this study, the composite score function is defined as:
\begin{equation}
S(\widehat{\boldsymbol{\theta}}) = w_1 R^2(\widehat{\boldsymbol{\theta}}) + w_2  C(\widehat{\boldsymbol{\theta}})\,, 
\label{eqn:composite_score}
\end{equation}
where $w_i \ge 0$ are user-specified weights such that $\sum_{i=1}^{2}w_i = 1$. The term $R^2(\boldsymbol{\widehat{\theta}})$ denotes the coefficient of determination used to evaluate model fit, calculated as 
\[
R^2(\widehat{\boldsymbol{\Theta}}) = 1 - \frac{\sum_{t=1}^{T} e_t^2}{\sum_{t=1}^{T} (Y_t-\bar{Y})^2}\ ,
\]
where $\bar{Y} = \frac{1}{T}\sum_{t=1}^{T}Y_t$ denotes the sample mean, $e_t = Y_t - \widehat{Y}_t$ is the model residual, and $\widehat{Y}_t = \left(\boldsymbol U_t^\top \widehat{\boldsymbol \beta}_U\right) \left(1 + \boldsymbol X_t(\widehat{\boldsymbol \alpha}, \widehat{\boldsymbol \gamma}, \widehat{\boldsymbol \kappa})^\top \widehat{\boldsymbol \beta}_X\right) + \boldsymbol Z_t^\top \widehat{\boldsymbol \beta}_Z$ is the fitted value. 

The term $C(\widehat{\boldsymbol{\theta}})$ assesses the constraint satisfaction, which basically counts the proportion of constraints satisfied by the parameter set. For example, if all constraints are satisfied, then $C(\widehat{\boldsymbol{\theta}})=1$. In our case, the following parameter constraints are adopted for numerical stability:
\begin{eqnarray}
\left\{
\begin{array}{l}
-\infty \le \beta_{U,i} \le \infty \ \ \ \mbox{ for } i=1,\ldots,p_U, \\
-\infty \le \beta_{X,m} \le \infty \ \ \ \mbox{ for } m=1,\ldots,M, \\
-\infty \le \beta_{Z,i} \le \infty \ \ \ \mbox{ for } i=1,\ldots,p_Z, \\
0 \le \alpha_m \le 1 \ \ \ \mbox{ for } m=1,\ldots,M,  \\
1 \le \gamma_m \le 5 \ \ \ \mbox{ for } m=1,\ldots,M, \\
0.5 \le \kappa_m \le 5 \ \ \ \mbox{ for } m=1,\ldots,M. \\
\end{array}
\right.
\label{eqn:M2_constraints}
     \end{eqnarray}

By default, weights are set equally ($w_i = 1/2$), but can be adjusted based on specific modeling priorities. For instance, if model fit is a priority, then set $w_1>w_2$, say, setting $w_1=0.75$ gives model fit three times as much priority. We also note that (\ref{eqn:composite_score}) can be easily customized to include additional components if required; this will be revisited in Section \ref{sec:econ}.

\subsection{Retaining elites}
\label{sec:elites}

To ensure the efficiency of our proposed GA, an evolutionary selection mechanism is incorporated to mimic the natural selection process in biological evolution, whereby high-performing candidates (referred to as ``elites'') are retained while less competitive candidates are discarded.

In each iteration, the entire population is evaluated using the multi-objective score defined in (\ref{eqn:composite_score}). The top-performers, specifically the first $n_e=1,000$ members, are retained as elites and propagate to the next iteration. The rest is discarded. Repetitively doing this trims out poor candidates and routinely stores potential candidates, resulting in monotonically increasing scores per iteration, enhancing the procedure.

\subsection{The Complete GA}

Finally, we combine all the processes described in Section \ref{sec:init}--\ref{sec:elites} to construct the complete GA, as summarized below: 
\begin{enumerate}
    \item Start with an initial value, $\boldsymbol{\theta}^{(0)}$.
    \item Create an initial population of size $n_0$,
    \[
\mathbf{P}^{(0)}
=
\{\boldsymbol{\theta}^{(1)},\ldots,\boldsymbol{\theta}^{(n_0)}\},
\]
as described in Section \ref{sec:init} with $a=0.5$, $b=2.5$, and $\sigma_{\text{init}}=0.5$.
    \item Set $i=0$ so that the current population is denoted as $\mathbf{P}^{(i)}$ with size $n_i$.
    \item Members of the current population $\mathbf{P}^{(i)}$ undergo breeding and mutation for $g=5$ generations. In each generation, the numbers of offspring and mutations are given by $n_i r_b$ and $n_i r_m$, respectively.
    \begin{itemize}
        \item Breed: For two randomly selected population members $\boldsymbol{\theta}^{(1)}$ and $\boldsymbol{\theta}^{(2)}$, and for $j=1,\ldots,p$,
        \[
        \begin{cases}
            \text{Direct inheritance: } & \theta_j^{(\text{child})} = \left\{\begin{array}{ll}
     \theta_j^{(1)}, & \mbox{ with probability 0.25,} \\
     \theta_j^{(2)}, & \mbox{ with probability 0.25.} \\
    \end{array}\right. \\
            \text{Arithmetic crossover: } & 
\theta_j^{(\text{child})}
=
\pi \theta_j^{(1)}
+
(1-\pi)\theta_j^{(2)}, \mbox{where }
\pi \sim U(0,1), \mbox{ with probability 0.25.}
\\
            \text{BLX-$\alpha$ crossover: } & 
\theta_j^{(\text{child})}
\sim
U\left(
\min\left\{\theta_j^{(1)},\theta_j^{(2)}\right\} - \delta d_j,\,
\max\left\{\theta_j^{(1)},\theta_j^{(2)}\right\} + \delta d_j
\right), \\
 & \mbox{with probability 0.25, where $\delta = 0.30$.}
        \end{cases}
        \]

        \item Mutate: For a randomly selected population member $\boldsymbol{\theta}$, for each $j=1,\ldots,p$,
        \[
        \begin{cases}
            \text{Gaussian mutation: With probability 0.50, } & \theta_j \leftarrow \theta_j + \varepsilon_j,
\qquad
\varepsilon_{j,\text{mutate}} \sim N(0,\sigma_{j,\text{mutate}}^2)\ , \\
            \text{No mutation: With probability 0.45, } & \theta_j\leftarrow \theta_j\ , \\
            \text{Sign-flip mutation: With probability 0.05, } & \theta_j\leftarrow -\theta_j\ , \\
        \end{cases}
        \]
        where we set $\sigma_{j,\text{mutate}}=1$ for our analysis.
    \end{itemize}
    \item Evaluate the population members using the scoring metric in (\ref{eqn:composite_score}).
    \item Rank the population members according to their scores and retain $n_e$ members with highest scores as the elites,
        \[
\mathbf{P}^{(i+1)}
=
\{\boldsymbol{\theta}^{(1)},\ldots,\boldsymbol{\theta}^{(n_e)}\}.
\]
    \item Set $i=i+1$ and repeat for $b-1$ times the steps 4-6.
    \item The result of the above is $\mathbf{P}^{(0)}, \ldots, \mathbf{P}^{(b)}$, the solution of which is the member of $\mathbf{P}^{(b)}$ with the highest score.
\end{enumerate}
A summary of the algorithm is presented in Algorithm \ref{alg:ga}.

\begin{algorithm}[H]
\caption{Genetic Optimization Procedure}
\label{alg:ga}
\begin{algorithmic}[1]

\State Initialize parameter vector $\boldsymbol{\theta}^{(0)}$.

\State Generate an initial population
\[
\mathbf{P}^{(0)}=\{\boldsymbol{\theta}^{(1)},\ldots,\boldsymbol{\theta}^{(n_0)}\}.
\]

\For{$i=0,\ldots,b-1$}

\For{$g=1,\ldots,5$}

    \State Generate $n_i r_b$ offspring from randomly selected parent pairs $\boldsymbol{\theta}^{(1)}$ and $\boldsymbol{\theta}^{(2)}$ from $\mathbf{P}^{(i)}$ of size $n_i$: 
    
    \For{each parameter $j=1,\ldots,p$}

       one of:

        \begin{itemize}
            \item Direct inheritance from parent 1, $\theta_j^{(\text{child})}=\theta_j^{(1)}$ (probability $0.25$);
            \item Direct inheritance from parent 2, $\theta_j^{(\text{child})}=\theta_j^{(2)}$ (probability $0.25$);
            \item Arithmetic crossover (probability $0.25$)
            \[
            \theta_j^{(\text{child})}
            =
            \pi\theta_j^{(1)}
            +(1-\pi)\theta_j^{(2)},
            \qquad
            \pi\sim U(0,1) \ ;
            \]
            \item BLX-$\alpha$ crossover with $\delta=0.30$
            (probability $0.25$)
            \[
            \theta_j^{(\text{child})}
\sim
U\left(
\min\left\{\theta_j^{(1)},\theta_j^{(2)}\right\} - \delta d_j,\,
\max\left\{\theta_j^{(1)},\theta_j^{(2)}\right\} + \delta d_j
\right).
            \]
        \end{itemize}

    \EndFor
    \State Apply mutations to $n_i r_m$ offspring:
    \For{each parameter $j=1,\ldots,p$}

    one of:

    \begin{itemize}
        \item Gaussian mutation $\theta_j \leftarrow N(\theta_j,\sigma_{j,\text{mutate}}^2)$, (probability $0.50$);
        \item No mutation $\theta_j \leftarrow \theta_j$ (probability $0.45$);
        \item Sign-flip mutation $\theta_j \leftarrow -\theta_j$ (probability $0.05$).
    \end{itemize}

    \EndFor

\EndFor

\State Evaluate all candidates in the current population using the score function in (\ref{eqn:composite_score}).

\State Rank candidates and retain the top $n_e$ elites:
\[
\mathbf{P}^{(i+1)}
=
\{\boldsymbol{\theta}^{(1)},\ldots,\boldsymbol{\theta}^{(n_e)}\}.
\]

\EndFor

\State Return the highest-scoring member of $\mathbf{P}^{(b)}$.

\end{algorithmic}
\end{algorithm}

\subsection{Genetic optimization results}

We now evaluate the use of the GA for estimating the proposed MMM. Table \ref{tab:genetic_tab_result} presents the estimated parameters, $\boldsymbol{\widehat{\theta}}$ and $\widehat{\sigma}$, using various algorithms in a single run for illustrative purposes. Also included as a measure of overall fitness is the Mean Absolute Error (MAE) calculated as
\begin{equation}
    \operatorname{MAE} = \frac{1}{Rp}\times\sum_{i=1}^{R}\sum_{j=1}^{p}\left|\theta_j-\widehat{\theta}^{(i)}_j\right|, 
\end{equation}
and the Mean Absolute Percentage Error (MAPE) calculated as
\begin{equation}
    \operatorname{MAPE} = \frac{1}{Rp}\times\sum_{i=1}^{R}\sum_{j=1}^{p}\left|\frac{\theta_j-\widehat{\theta}^{(i)}_j}{\theta_j}\right|, 
\end{equation}
where $\widehat{\theta}^{(i)}_j$ is the estimate of $\theta_j$ in the $i^{th}$ run.

\begin{table}[ht]
\caption{Estimated parameters of M2 using MLE, LSE, and GAs.}
\label{tab:genetic_tab_result}
\centering
\small
\begin{tabular}{rrrrr}
  \hline
 & TRUE & MLE & LSE & GA \\ 
  \hline
  $\beta_{U,1}$ & -1.00000 & -1.67246 & -1.59125 & -0.87711 \\ 
  $\beta_{U,2}$ & 1.00000 & 1.82877 & 1.74061 & 0.97588 \\ 
  $\beta_{U,3}$ & 0.50000 & 0.67560 & 0.64385 & 0.35399 \\ 
  $\beta_{U,4}$ & 2.00000 & 3.48460 & 3.31818 & 1.85441 \\ 
  $\beta_{U,5}$ & 5.00000 & 8.80858 & 8.38258 & 4.53991 \\ 
  $\beta_{U,6}$ & -3.00000 & -5.28647 & -5.03079 & -2.74259 \\ 
  $\beta_{X,1}$ & -0.50000 & -8.83319 & -7.70470 & 0.19520 \\ 
  $\beta_{X,2}$ & 1.00000 & 9.15313 & 8.11766 & 3.03335 \\ 
  $\beta_{X,3}$ & 2.00000 & 0.472088 & 0.471395& -0.41447 \\ 
  $\alpha_1$ & 0.50000 & 0.343933 & 0.375217 & 0.078566 \\ 
  $\alpha_2$ & 0.50000 & 0.529501 & 0.528544 & 0.530248 \\ 
  $\alpha_3$ & 0.50000 & 0.474554 & 0.467816 & 0.907275 \\ 
  $\gamma_1$ & 3.00000 & 0.714352 & 0.948086 & 3.82956 \\ 
  $\gamma_2$ & 3.00000 & 0.958592 & 0.954822 & 3.535059 \\ 
  $\gamma_3$ & 3.00000 & 5.114809 & 5.152541 & 3.116623 \\ 
  $\kappa_1$ & 1.00000 & 0.002363 & 0.011043 & 4.869494 \\ 
  $\kappa_2$ & 1.00000 & 0.044866 & 0.053470 & 0.792407 \\ 
  $\kappa_3$ & 1.00000 & 1.217172 & 1.21867 & 1.515492 \\ 
  $\beta_{Z,1}$ & -1.00000 & -1.29190 & -1.29142 & -0.95995 \\ 
  $\beta_{Z,2}$ & 0.50000 & 0.38810 & 0.38874 & 0.60006 \\ 
  $\beta_{Z,3}$ & 1.00000 & 1.00121 & 1.00061 & 1.16838 \\ 
  $\beta_{Z,4}$ & 6.00000 & 5.87288 & 5.87374 & 5.91421 \\ 
  $\sigma$ & 5.00000 & 4.948082 & 5.002961 & 5.509203 \\ 
   \hline
   MAD &  & 1.59468 & 1.44261 & 0.61458 \\
   \hline
\end{tabular}
\end{table}

\begin{table}[ht]
\caption{The MAE and MAPE for the estimated parameters of M1 and M2 configurations using MLE, LSE, and GAs.}
\label{tab:MAE_MAPE_result}
\centering
\small
\begin{tabular}{rrrrrr}
  \hline
 &  & MLE & LSE & GA \\ 
   \hline
   M1 &  MAE  & $0.083734$ & $0.083581$ & $ 0.090151$ \\
 &  MAPE  & $0.062933$ & $0.062906$ & $0.078991$ \\
 \hline
   M2 &  MAE  & $16.270440$ & $21.086771$ & $1.233041$ \\
    &  MAPE  & $6.578072$ & $7.801424$ & $0.802556$ \\
   \hline
\end{tabular}
\end{table}

\begin{figure}[htbp]
\centering
\includegraphics[scale=0.5]{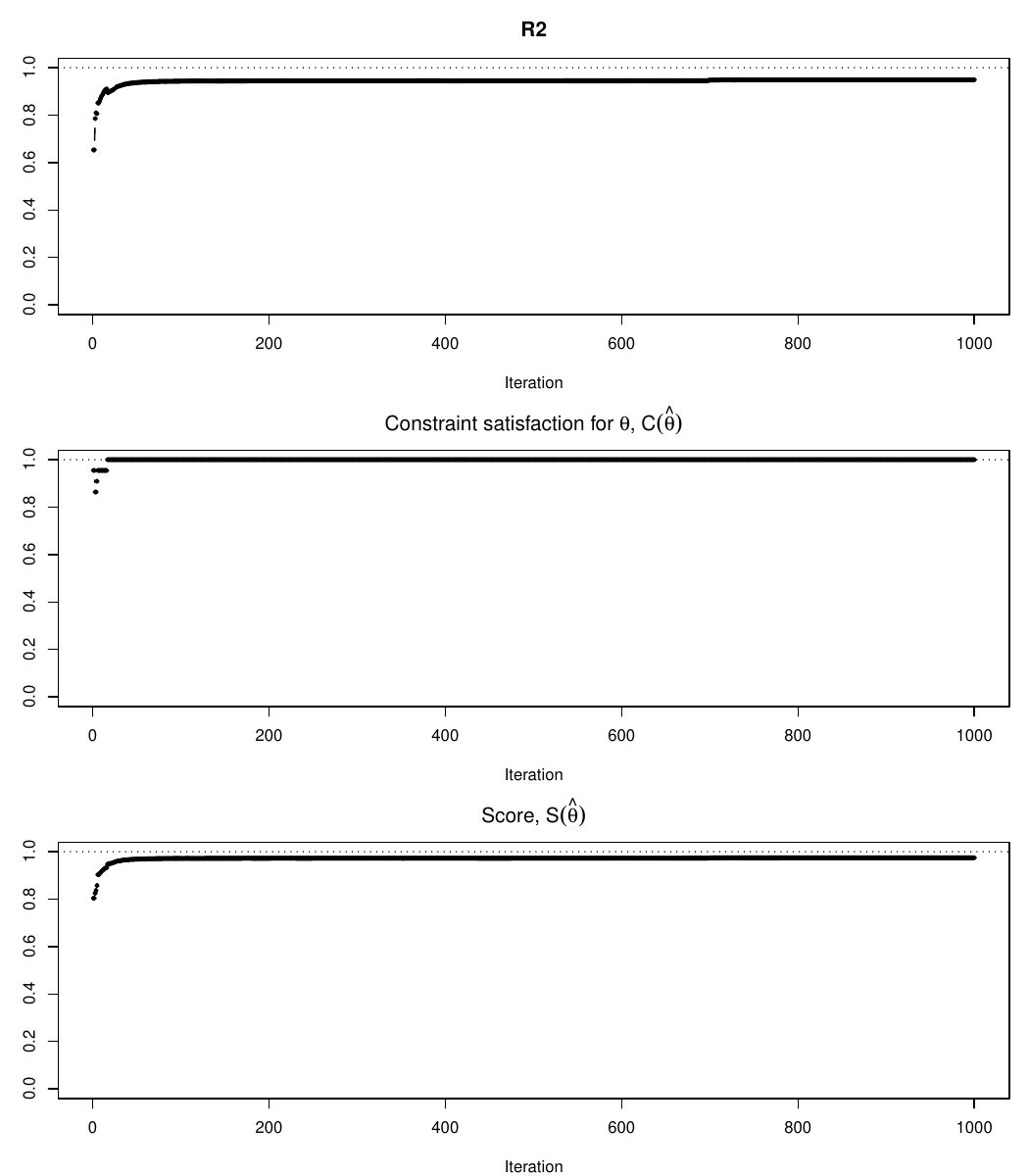}
\caption{\label{fig:genetic_score_plot_result}Plot of the $R^2$ values, constraint satisfactions, $C(\boldsymbol{\widehat{\theta}})$, and the composite scores (as defined in \ref{eqn:composite_score}) against iteration.}
\end{figure}

Figure \ref{fig:genetic_score_plot_result} illustrates the trajectory of the GA during the search process. As expected, the scores $S(\widehat{\boldsymbol{\theta}})$ exhibit an upward trend, indicating that the algorithm progressively explores combinations of $\boldsymbol{\theta}$ associated with higher objective values. This monotonic improvement is a direct consequence of the elitist selection mechanism adopted in the proposed GA. The trajectory also suggests signs of convergence, although the final score of $0.9744480$ indicates that further improvement may still be possible. In particular, further investigation is needed to determine whether the algorithmic tuning parameters within the GA have the potential to enhance estimation accuracy beyond their influence on the rate of convergence.

It is worth noting that, while the $R^2$ values generally increase over the course of the optimization, they are not necessarily monotonic as the overall score. This is because the GA optimizes the composite objective function rather than model fit alone, and may therefore prioritize improvements in constraint satisfaction over marginal changes in $R^2$ when doing so leads to a higher overall score. This behavior highlights an important advantage of the GA framework, which enables simultaneous optimization of model fit and constraint satisfaction without prematurely excluding solutions that temporarily violate such constraints. This facilitates exploration of infeasible regions while searching for high-quality feasible solutions.

Table \ref{tab:MAE_MAPE_result} presents the MAE and MAPE computed under M1 and M2 configurations for quantitative comparison. Under M1, the performances of all approaches are similar across the board, with the LSE approach having a slight edge. Specifically, both the MAE and MAPE for GA are slightly larger than those for conventional approaches. This implies that the strength of the GA is not apparent in this setting because the problem is relatively straightforward without the need for a more heuristic approach. Under M2, both conventional optimization approaches struggle to accurately estimate the parameters due to the identifiability issues described in Section \ref{sec:identifiability}. In contrast, the proposed GA outperforms its counterpart considerably by yielding much smaller MAE and MAPE. The advantage of adopting the GA becomes apparent due to its ability to maintain numerical stability through accounting for constraints flexibly without boundary issues. This allows the incorporation of nonlinear relationships in media-related variables within the MMM framework by facilitating robust parameter estimation. 

\section{Economical constraints}
\label{sec:econ}

In addition to imposing parameter constraints, it is practically useful to consider budget constraints that govern how resources can be allocated across marketing channels. They can consist of collective contributions respectively due to underlying demand (baseline), media activities, and external shocks, which should be dynamically bounded to ensure commercial realism and statistical validity in a corporate environment. This ensures that the influence attributed to temporal patterns, economic conditions, pricing, distribution, exogenous factors, and brand metrics remains within empirically and commercially reasonable bounds. The factors that inform these bounds are derived from a qualitative and quantitative understanding of market dynamics, reflecting insights often gained from market research, business intelligence, and industry benchmarks.

Mathematically, we impose for $t=1,\ldots,T$,
\begin{equation}
    \left.\begin{array}{ll}
      \phi_U^{\text{min}}\le \boldsymbol U_t^\top \boldsymbol \beta_U \le \phi_U^{\text{max}} & \mbox{ for total baseline/underlying demand contribution,} \\
      \phi_X^{\text{min}}\le \boldsymbol X_t^\top \boldsymbol \beta_X \le \phi_X^{\text{max}} & \mbox{ for total media contribution,} \\
      \phi_Z^{\text{min}}\le \boldsymbol Z_t^\top \boldsymbol \beta_Z \le \phi_Z^{\text{max}} & \mbox{ for total external shocks contribution,} \\
    \end{array}
    \right\}
    \label{eqn:contrib}
\end{equation}
where $\phi_{\cdot}^{\text{min}}$ and $\phi_{\cdot}^{\text{max}}$ are the lower and upper bounds respectively. For example, we could impose that the total media contribution should exceed $10\%$ but should not dominate more than $80\%$ of the total KPI, which we would then set $\phi_{X}^{\text{min}}=0.1Y_t$ and $\phi_{X}^{\text{max}}=0.8Y_t$. 

In the conventional optimization framework, accounting for these contribution-specific constraints may not be straightforward and may cause further estimation issues, e.g. numerical instability due to boundary issues. On the other hand, the GA allows incorporation of such constraints naturally through the built-in scoring system. To achieve this, the composite score function in (\ref{eqn:composite_score}) can be modified accordingly as follows:
\begin{equation}
S(\widehat{\boldsymbol{\theta}}) = w_1 R^2(\widehat{\boldsymbol{\theta}}) + w_2  C(\widehat{\boldsymbol{\theta}})+ w_3  C_U(\widehat{\boldsymbol{\theta}})+ w_4  C_X(\widehat{\boldsymbol{\theta}})+ w_5  C_Z(\widehat{\boldsymbol{\theta}})\,, 
\label{eqn:composite_score2}
\end{equation}
where $\sum_{i=1}^{5}w_i = 1$. Similarly to before, the terms $C_U(\widehat{\boldsymbol{\theta}})$, $C_X(\widehat{\boldsymbol{\theta}})$, and $C_Z(\widehat{\boldsymbol{\theta}})$ assess the constraint satisfactions, which basically count the proportion of constraints satisfied by the baseline, media, and external shocks contributions expressed in (\ref{eqn:contrib}). For example, if all constraints are satisfied, then $C_U(\widehat{\boldsymbol{\theta}})=C_X(\widehat{\boldsymbol{\theta}})=C_Z(\widehat{\boldsymbol{\theta}})=1$. By default, weights are set equally ($w_i = 1/5$), but can again be adjusted based on specific modeling priorities. We remark that the resulting solution might be slightly different from the true values used to simulate the data, but it provides customizations that allow practitioners to better calibrate resource allocation according to their own business needs, while adhering closely to the actual solution.

\subsection{Sensitivity analysis}

\begin{figure}[htbp]
\centering
\includegraphics[scale=0.5]{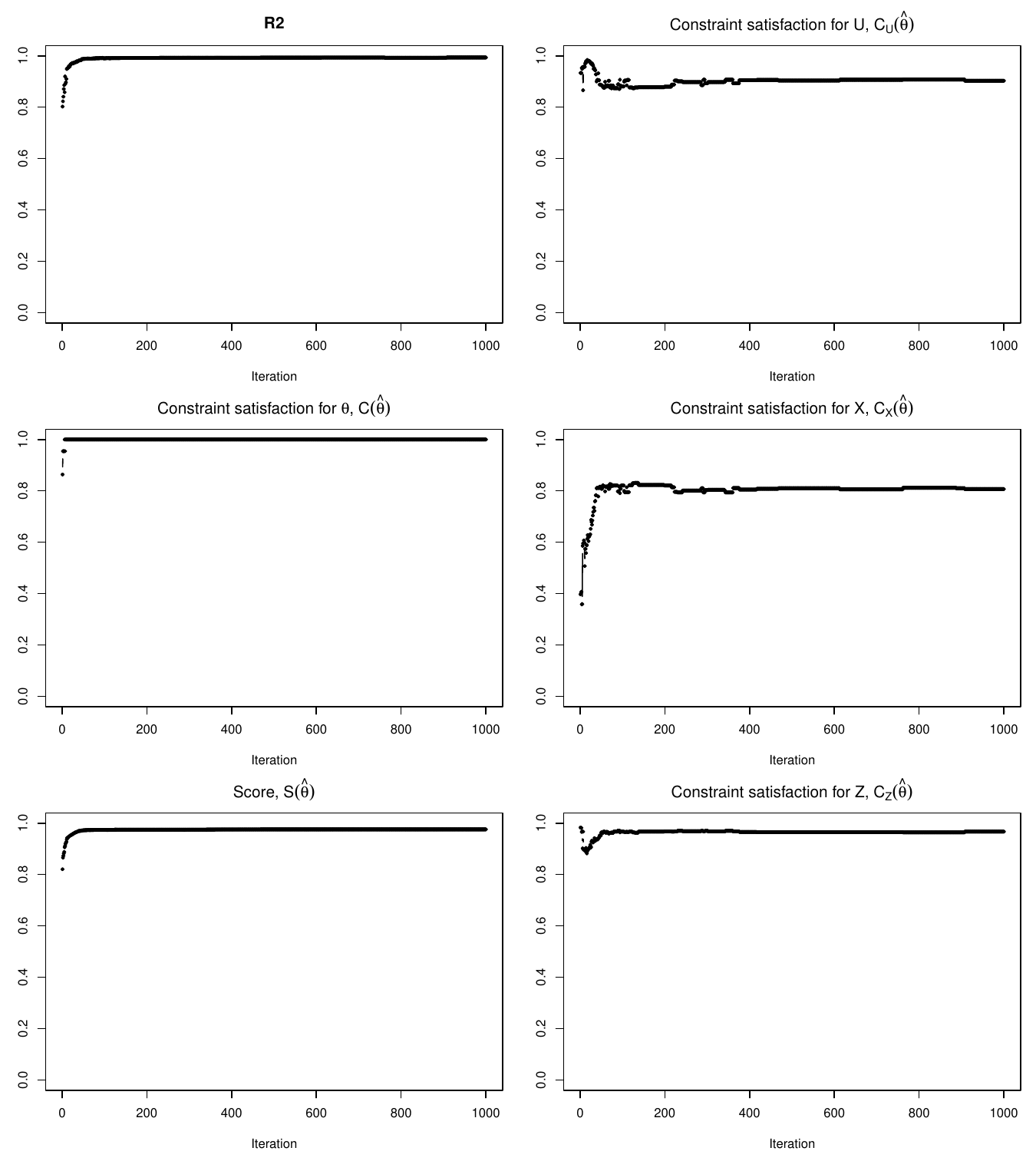}
\caption{\label{fig:genetic_score_modified_plot_result}Plot of the $R^2$ values, the composite scores (as defined in \ref{eqn:composite_score}), and all other constraint satisfactions against iteration, when the weightings are $\{0.35,0.35,0.1,0.1,0.1\}$.}
\end{figure}

We conduct a sensitivity analysis to assess the impact of different weight combinations $w_i$ $(i=1,\dots,5)$ on the estimation procedure. Specifically, the simulation study in Section \ref{sec:sim_study} is repeated under a modified experimental design to intended to better reflect practical settings. For illustrative purposes, we set
\begin{equation}
        \left.\begin{array}{ll}
      \phi_U^{\text{min}} = 0.05Y_t, \phi_U^{\text{max}} = 0.30Y_t & \mbox{ for total baseline contribution,} \\
      \phi_X^{\text{min}} = 0.005Y_t, \phi_X^{\text{max}} = 0.65Y_t & \mbox{ for total media contribution,} \\
      \phi_Z^{\text{min}} = 0.001Y_t, \phi_Z^{\text{max}} = 0.70Y_t & \mbox{ for total external shocks contribution.} \\
    \end{array}
    \right\}
    \label{eqn:bounds}
\end{equation}
We then consider $w_i$ $(i=1,\dots,5)$ from the following set: 
\begin{itemize}
\item Case 1: $\{0.5,0.5,0,0,0\}$;
\item Case 2: $\{0.4,0.4,1/15,1/15,1/15\}$;
\item Case 3: $\{0.35,0.35,0.1,0.1,0.1\}$;
\item Case 4: $\{0.2,0.2,0.2,0.2,0.2\}$.
\end{itemize}
So from Case 1 to 4, we put increasingly more priorities on the contribution-specific constraints. As an illustration, Figure \ref{fig:genetic_score_modified_plot_result} depicts the evolution of each component of the composite score (\ref{eqn:composite_score2}) for Case 3. 

The resulting estimated parameters are presented in Table \ref{tab:genetic_sensitivity_analysis_result}. From Case 1 to 4, as more priority is given to contribution-specific constraints, the more estimated parameters deviate away from the true values. However, we remark that this is expected as the solution is intended to reflect higher contribution-specific constraint satisfaction rate for business needs, while adhering closely to the true values. 

\begin{table}[ht]
\caption{Sensitivity analysis with respect to various cases of weight combinations $\{w_1,w_2,w_3,w_4,w_5\}$ for the modified GA.}
\label{tab:genetic_sensitivity_analysis_result}
\centering
\small
\begin{tabular}{rrrrrr}
  \hline
 & TRUE & Case 1 & Case 2 & Case 3 & Case 4 \\
  \hline
  $\beta_{U,1}$ & -0.10000 & -0.139247 & -0.107760 & -0.067141 & -0.391967 \\ 
  $\beta_{U,2}$ & 1.00000 & 1.207156 & 1.773270 & 1.741407 & 1.741174 \\ 
  $\beta_{U,3}$ & 0.50000 & 0.633257 & 0.636381 & 0.618247 & 0.879214 \\  
  $\beta_{U,4}$ & 2.00000 & 2.400596 & 3.913582 & 3.911433 & 3.959423 \\ 
  $\beta_{U,5}$ & 6.00000 & 7.218087 & 12.083312 & 12.085589 & 11.900188 \\ 
  $\beta_{U,6}$ & -0.30000 & -0.379780 & -0.756956 & -0.731628 & -0.819350 \\  
  $\beta_{X,1}$ & -0.50000 & 0.406627 & -1.017928 & 0.318347 & 0.400395 \\  
  $\beta_{X,2}$ & 5.00000 & 3.602452 & 2.434086 & 2.038239 & 1.738707 \\ 
  $\beta_{X,3}$ & 2.00000 & 1.260951 & 1.285321 & 0.590078 & 0.877872 \\ 
  $\alpha_1$ & 0.50000 & 0.919742 & 0.773775 & 0.002918 & 0.222300 \\ 
  $\alpha_2$ & 0.50000 & 0.493118 & 0.489232 & 0.514804 & 0.516093 \\ 
  $\alpha_3$ & 0.50000 & 0.496556 & 0.533190 & 0.619294 & 0.544609 \\ 
  $\gamma_1$ & 3.00000 & 2.875556 & 3.626780 & 2.869855 & 2.571700 \\ 
  $\gamma_2$ & 3.00000 & 3.330582 & 3.051579 & 3.503814 & 3.317199 \\ 
  $\gamma_3$ & 3.00000 & 3.890807 & 2.338578 & 3.380626 & 2.481749 \\ 
  $\kappa_1$ & 1.00000 & 1.447208 & 1.636440 & 4.648126 & 4.507199 \\ 
  $\kappa_2$ & 1.00000 & 1.090657 & 1.000452 & 1.210528 & 1.324303 \\ 
  $\kappa_3$ & 1.00000 & 1.119885 & 0.956723 & 1.801160 & 1.003541 \\ 
  $\beta_{Z,1}$ & -0.10000 & 0.289785 & 1.718349 & 1.693682 & 2.422461 \\ 
  $\beta_{Z,2}$ & 0.50000 & -0.206471 & 2.027822 & 2.179433 & 2.536504 \\ 
  $\beta_{Z,3}$ & 1.00000 & 1.115168 & 2.697521 & 2.606189 & 2.928687 \\ 
  $\beta_{Z,4}$ & 3.00000 & 2.989461 & 4.004820 & 4.122444 & 3.846598 \\ 
  $\sigma$ & 1.00000 & 1.135891 & 2.124362 & 2.266448 & 2.461206 \\ 
   \hline
   MAD &  & 0.3875 & 0.9861 & 1.2298 & 1.2743 \\ 
   \hline
\end{tabular}
\end{table}

\section{Empirical analysis}
\label{sec:empirical}

The company under study employs a comprehensive marketing strategy encompassing nine distinct media channels: Marketplace, Display, Email, Paid Search, Paid Social, Organic, Out-of-Home (OOH), Public Relations (PR), and TV Broadcast. The analysis spans the period from 23 October 2022 to 16 November 2025, providing a dataset\footnote{Note that the data has been anonymized to ensure compliance with privacy and data-protection standards.} of dimension $T \times K$, where $T=160$ weeks and $K=12$ variables (which is not necessarily equal to the number of parameters, $p$) for parameter estimation. The average weekly KPI ($Y_t$) for the company during this period was approximately $\pounds$287,931.63. The underlying demands $U_t$ considered in the study include seasonal effects (monthly), yearly effect, promotions, economic variables, and customer traffic. In this study, no external shock variable $Z_t$ is considered. Information on the constraints associated with the underlying demands and the media-specific parameters is provided in the Appendix. 

\subsection{Optimization using GA}

In this particular empirical analysis, we propose the following multi-objective score function:
\begin{equation}\label{eqn:score_function2}
S(\widehat{\boldsymbol{\theta}}) = w_1 R^2(\widehat{\boldsymbol{\theta}}) + w_2  C(\widehat{\boldsymbol{\theta}})+ w_3  C_U(\widehat{\boldsymbol{\theta}})+ w_4  C_X(\widehat{\boldsymbol{\theta}}) + w_5 (1 - |DW(\widehat{\boldsymbol{\theta}}) - 2|)\,,
\end{equation}
where 
\begin{equation*}
DW(\widehat{\boldsymbol{\theta}}) = \frac{\sum_{t=2}^{T}(e_t-e_{t-1})^2}{\sum_{t=1}^{T}e_t^2}
\end{equation*}
is the Durbin-Watson (DW) statistic. In this study, weights are set equally ($w_i = 1/5$) such that $\sum_{i=1}^{5}w_i = 1$. Higher scores denote better-performing models. 

Note that the score function proposed in \eqref{eqn:score_function2} for this study is analogous to \eqref{eqn:composite_score2}, where the $R^2$ component quantifies the proportion of variance explained by the model and serves as a measure of overall goodness of fit, while the constraint satisfaction components guide the parameter search towards solutions that satisfy commercially meaningful requirements. The DW statistic is additionally tracked, aiming for a value close to 2.0 to indicate minimal autocorrelation in residuals. 

\begin{figure}[H]
    \centering
    \includegraphics[width=0.5\textwidth]{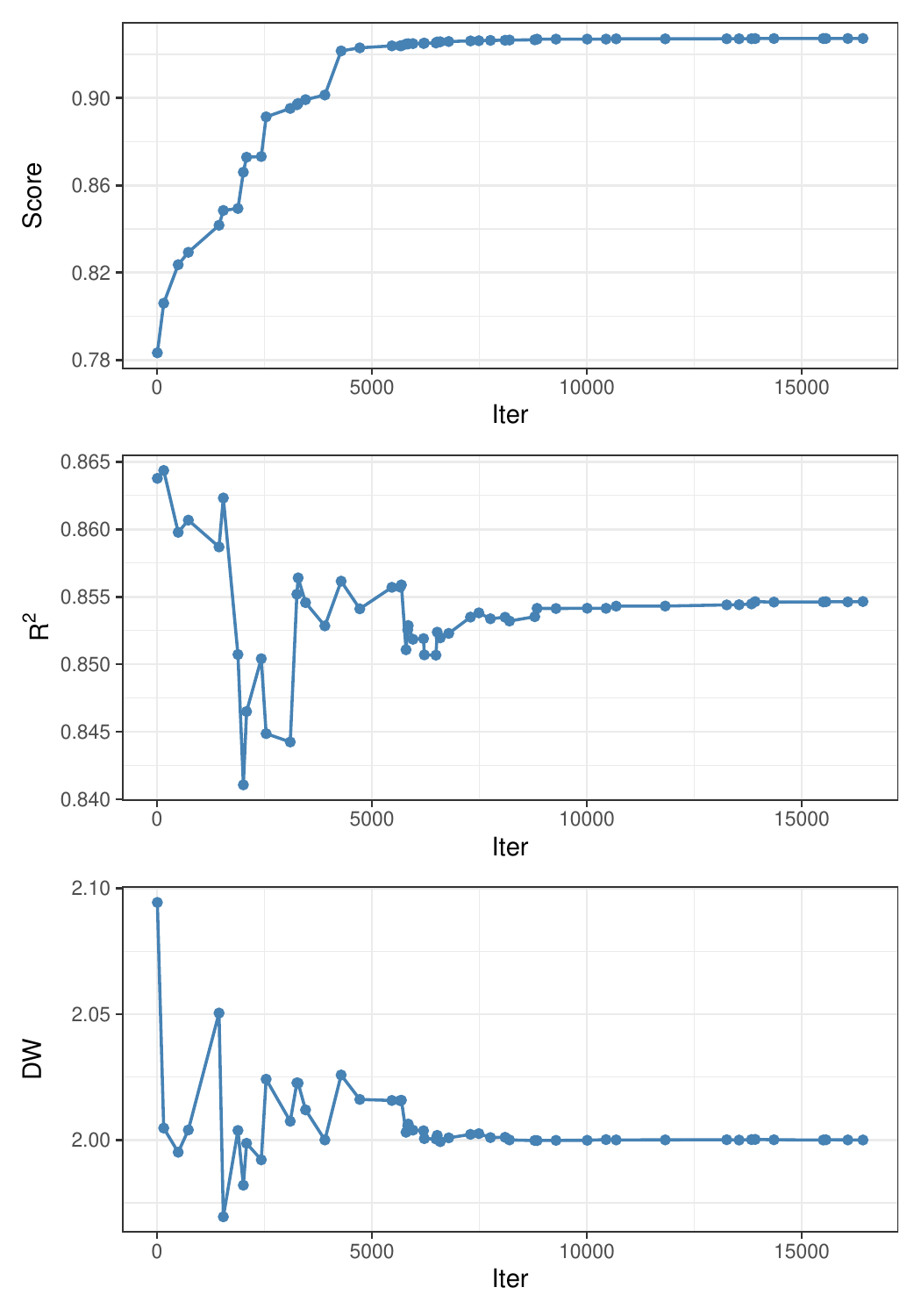} 
    \caption{Plot of the composite scores (as defined in (\ref{eqn:score_function2})), R2, and DW against iterations.}
    \label{fig:plot_eval_score}
\end{figure}

The GA is then implemented with \eqref{eqn:score_function2} to find the optimal solution. Table \ref{tab:table_gen_opt} illustrates the GA in action, where each row corresponds to an evaluation step in the optimization process, showing how the algorithm balances maximizing the model’s explanatory power with satisfying predefined constraints (the constraint satisfaction values are not reported in the table due to the proprietary nature of the underlying commercial information). The core performance measure is captured in the \textit{Score} column, which represents the composite score function defined in \eqref{eqn:score_function2}. Figure \ref{fig:plot_eval_score} also illustrates the progression of the GA in the search for a solution. Evidently, the composite score shows a consistent upward trend, demonstrating the algorithm’s effective progression toward improved model specifications. 

Overall, the estimation algorithm iteratively updates the model parameters, leading to progressive improvements in key metrics across iterations. In parallel, $R^2$ value stabilizes at approximately 0.855, while the DW statistic converges to around $2$ after the iteration count exceeds $10,000$. More specifically, the $R^2$ values generally increases with higher composite scores, confirming that optimization leads to models with stronger predictive power. For instance, $R^2$ improves from approximately $0.850670$ in earlier iterations to $0.854638$ at convergence, indicating that a substantial proportion of the variance in transaction volume is explained by the model. Table \ref{tab:table_gen_opt} also shows DW values consistently around 2.0, suggesting the absence of significant autocorrelation and confirming that the model captures temporal dependencies adequately. It is worth noting the GA does not simply pick the candidate with the highest $R^2$, but rather balances overall model fit, low residual autocorrelation, and high levels of constraint satisfaction. For example, \textit{Iter} 155 attains the highest $R^2$, but its score is penalized due to a poor DW statistic and low constraint satisfaction. By contrast, although \textit{Iter} 16424 does not yield the highest $R^2$, it achieves a stronger DW value and high constraint satisfaction, resulting in the highest score and is, therefore, selected as the final model.

\begin{table}[H]
\centering
\small
\caption{Evaluation results across Genetic Algorithm runs (highest to lowest score).}
\begin{tabular}{ccccc}
\hline
\textbf{\textit{Index}} & \textbf{\textit{Iter}} & \textbf{\textit{Score}} & $\boldsymbol{R^2}$ & $\boldsymbol{DW}$ \\
\hline
52 & 16424 & 0.927313 & 0.854638 & 1.999975 \\
51 & 16072 & 0.927306 & 0.854615 & 1.999994 \\
50 & 15562 & 0.927300 & 0.854627 & 2.000053 \\
49 & 15500 & 0.927294 & 0.854612 & 1.999950 \\
48 & 14353 & 0.927285 & 0.854599 & 2.000059 \\
47 & 13917 & 0.927266 & 0.854632 & 2.000199 \\
46 & 13831 & 0.927199 & 0.854464 & 2.000131 \\
45 & 13542 & 0.927186 & 0.854403 & 1.999936 \\
44 & 13254 & 0.927174 & 0.854389 & 2.000084 \\
43 & 11824 & 0.927149 & 0.854304 & 2.000011 \\
42 & 10682 & 0.927148 & 0.854297 & 2.000002 \\
41 & 10446 & 0.927037 & 0.854136 & 2.000125 \\
40 & 10008 & 0.927033 & 0.854142 & 1.999846 \\
39 & 9282  & 0.927019 & 0.854123 & 1.999827 \\
38 & 8842  & 0.927005 & 0.854134 & 1.999753 \\
37 & 8788  & 0.926699 & 0.853517 & 1.999761 \\
36 & 8200  & 0.926597 & 0.853194 & 1.999999 \\
35 & 8097  & 0.926478 & 0.853477 & 2.001044 \\
34 & 7757  & 0.926449 & 0.853365 & 2.000933 \\
33 & 7488  & 0.926276 & 0.853801 & 2.002500 \\
32 & 7293  & 0.926191 & 0.853485 & 2.002208 \\
31 & 6786  & 0.925933 & 0.852278 & 2.000823 \\
30 & 6489  & 0.925243 & 0.850670 & 2.000367 \\
29 & 6589  & 0.925804 & 0.851950 & 1.999314 \\
28 & 6519  & 0.925731 & 0.852372 & 2.001817 \\
27 & 6219  & 0.925213 & 0.850687 & 2.000522 \\
26 & 6200  & 0.925034 & 0.851894 & 2.003652 \\
25 & 5954  & 0.924951 & 0.851845 & 2.003886 \\
24 & 5844  & 0.924853 & 0.852862 & 2.006310 \\
23 & 5832  & 0.924843 & 0.852542 & 2.005713 \\
22 & 5793  & 0.924773 & 0.851069 & 2.003046 \\
21 & 5685  & 0.923988 & 0.855871 & 2.015790 \\
20 & 5656  & 0.923946 & 0.855706 & 2.015629 \\
19 & 5463  & 0.923922 & 0.855690 & 2.015691 \\
18 & 4717  & 0.923026 & 0.854099 & 2.016094 \\
17 & 4283  & 0.921617 & 0.856146 & 2.025824 \\
16 & 3905  & 0.901413 & 0.852835 & 1.999983 \\
15 & 3455  & 0.899289 & 0.854558 & 2.011959 \\
14 & 3283  & 0.897549 & 0.856392 & 2.022586 \\
13 & 3253  & 0.896917 & 0.855182 & 2.022695 \\
12 & 3099  & 0.895265 & 0.844237 & 2.007415 \\
11 & 2538  & 0.891395 & 0.844856 & 2.024134 \\
10 & 2425  & 0.873223 & 0.850405 & 1.992081 \\
9 & 2083  & 0.872911 & 0.846493 & 1.998659 \\
8 & 2008  & 0.866031 & 0.841065 & 1.981995 \\
7 & 1882  & 0.849416 & 0.850717 & 2.003768 \\
6 & 1542  & 0.848510 & 0.862308 & 1.969425 \\
5 & 1442  & 0.841729 & 0.858681 & 2.050446 \\
4 & 728   & 0.829343 & 0.860667 & 2.003964 \\
3 & 492   & 0.823658 & 0.859765 & 1.995100 \\
2 & 155   & 0.805994 & 0.864346 & 2.004713 \\
1 & 7     & 0.783273 & 0.863760 & 2.094427 \\
\hline
\end{tabular}%
\label{tab:table_gen_opt}
\end{table}

\subsection{Estimated Parameters}

The calibrated model provides specific parameter estimates for each media channel, including coefficients, adstock rates, and inflection points, as illustrated in Table \ref{tab:estimated_params}. These parameters are crucial for understanding the nonlinear response of transactions to media activity.

\begin{table}[htbp]
\centering
\small
\caption{Parameter Estimates for the final model.}
\label{tab:estimated_params}
\begin{tabular}{lccccc}
\hline
\textbf{Variable} & $\beta$ & $\gamma$ & $\alpha$ & $\kappa$  \\
\hline
Marketplace        & 5.40E-04 & 3.214286 & 0.300000 & 5.65E+05 \\
Display            & 1.86E-03 & 2.785714 & 0.300000 & 1.00E+06  \\
Email              & 5.80E-04 & 3.214286 & 0.300000 & 1.57E+06  \\
Paid Search        & 4.01E-02 & 2.642857 & 0.300000 & 3.14E+07 \\
Paid Social        & 2.76E-02 & 2.785714 & 0.385714 & 3.06E+07 \\
Organic            & 5.20E-04 & 3.500000 & 0.300000 & 2.50E+05 \\
OOH                & 1.41E-02 & 3.500000 & 0.300000 & 2.14E+06 \\
PR                 & 1.65E-02 & 3.500000 & 0.342857 & 1.50E+07  \\
TV Broadcast       & 3.54E-02 & 3.071429 & 0.300000 & 3.93E+06 \\
Customer/Traffic   & 6.97E-07 & - & - & - \\
Economic Variables & 4.04E-02 & - & - & -  \\
Promotions         & 3.01E-03 & - & - & -  \\
Seas\_M01          & 3.59E-01 & - & - & - \\
Seas\_M02          & 6.21E-02 & - & - & -  \\
Seas\_M03          & 8.81E-02 & - & - & - \\
Seas\_M04          & 2.52E-01 & - & - & -  \\
Seas\_M05          & 3.87E-01 & - & -  \\
Seas\_M06          & 4.01E-02 & - & - & -  \\
Seas\_M07          & 5.38E-02 & - & - & - \\
Seas\_M08          & 4.02E-01 & - & - & - \\
Seas\_M09          & 4.56E-02 & - & - & -  \\
Seas\_M10          & 7.71E-02 & - & - & -  \\
Seas\_M11          & 4.54E-02 & - & - & -  \\
Seas\_M12          & 3.46E-01 & - & - & -  \\
Yearly             & 2.11E-06 & - & - & - \\
\hline
\end{tabular}
\end{table}

The coefficient $\beta$ represents the effect of each variable on transactions. For instance, Paid Search has an estimated coefficient of 0.04009, indicating its influence after accounting for carryover and saturation effects. The parameters $\gamma_m$ ($m=1,\ldots,M$), which are constrained to range between 2.5 and 3.5, represent the shape parameters of the Hill function, influencing the curve's concavity or S-shape. All estimated shape parameters for media channels are within the range of 2.5 to 3.5. The adstock parameters  $\alpha_m$ ($m=1,\ldots,M$) capture the carryover effects of media activity. All media channels show an adstock rate of 0.3, except for Paid Social and PR which are respectively 0.386 and 0.343, suggesting relatively longer-lasting carryover effects for these two channels. The inflection parameters $\kappa_m$ ($m=1,\ldots,M$) signify the point of diminishing returns. For example, Paid Search has an inflection point of approximately 31.43 million, meaning that half of its maximum effectiveness is achieved at this level of adstocked impressions. 

\begin{figure}[H]
    \centering
    \includegraphics[width=0.5\textwidth]{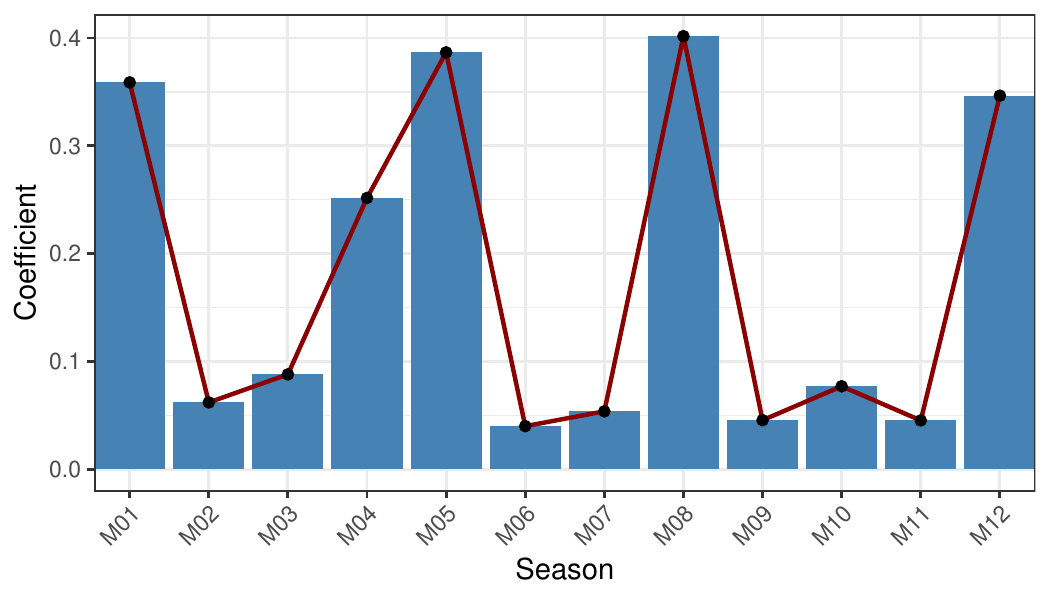}
    \caption{Coefficients of seasonal effects}
    \label{fig:seasonal}
\end{figure}

The estimated seasonal effects are illustrated in Figure \ref{fig:seasonal}. The results reveal a strong seasonal pattern, with notable peaks occurring during three periods of the year: the end-of-year holiday season (December--January), the Easter period (April--May), and the summer term (August). These findings provide actionable insights for improving marketing planning and resource allocation. For example, the results suggest that marketing activities may be strategically prioritized during these seasonal peaks to maximize their potential impact. As implied by the proposed MMM framework, this effect is further amplified through the multiplicative interaction between underlying demand and marketing activities, meaning that marketing investments made during periods of stronger underlying demand may generate greater incremental returns. Such effects can be quantified using the proposed model to support future marketing planning and allocation decisions.

\section{Conclusion}
\label{sec:conclusion}

This paper proposes an alternative MMM framework that explicitly incorporates a multiplicative interaction between baseline demand and media activities. Unlike conventional MMM specifications that assume additive effects, the proposed formulation allows marketing effectiveness to vary with the underlying level of consumer demand, thereby providing a more realistic representation of the incremental sales lift generated by advertising when purchasing intent is already strong.

The additional flexibility offered by the proposed model, however, comes at the cost of increased estimation complexity. Similar to conventional MMM frameworks, the nonlinear media transformations introduce weak identifiability of model parameters, while the multiplicative interaction further complicates the estimation. To address these challenges, we developed a highly-customizable multi-objective GA approach that performs optimization while naturally accommodating economically meaningful constraints. Both the simulation and empirical studies demonstrate that the proposed GA substantially improves the robustness and accuracy of parameter estimation relative to conventional optimization approaches. Moreover, the framework readily accommodates additional business rules and commercial constraints without requiring substantial modifications to the underlying estimation procedure.

Despite these advantages, several limitations remain. First, the iterative population-based nature of the GA results in higher computational costs than conventional local optimization methods. Second, although the proposed framework provides a practical and effective approach for navigating a highly non-convex optimization landscape, theoretical properties such as the uniqueness of the resulting solutions and convergence to the global optimum remain difficult to establish.

Several avenues for future research may further unlock the proposed approach's full potential. Improving the computational efficiency of the proposed algorithm through parallelization, adaptive search strategies, or hybrid optimization techniques would enhance its practical applicability to large-scale MMM problems. From a theoretical perspective, establishing convergence properties and quantifying convergence rates would provide a stronger foundation for evaluating and comparing alternative GA implementations. In addition, while a sensitivity analysis has been conducted to investigate the influence of the weighting parameters $w_i$ ($i=1,\ldots,5$) used in the composite scoring function, a theoretical understanding of how these weights affect the optimization landscape and the resulting parameter estimates remains an open problem. Finally, the proposed GA contains several algorithmic tuning parameters, including the initial population size ($n_0$), initialization perturbation ($\sigma_{\text{init}}^2$), breeding and mutation rates ($r_b$ and $r_m$), mutation variances ($\sigma_{j,\text{mutation}}^2$), elite population size ($n_e$), and the probabilities governing crossover and mutation operators, that were selected empirically in this study. Developing principled calibration or adaptive tuning strategies for these hyperparameters is another promising avenue for future research and may further improve both estimation accuracy and computational efficiency.

\section*{Acknowledgment}
The authors appreciate the comments from the editor, associate editor, and all reviewers involved, who helped improve the quality of the paper. 

\section*{Funding}
This research was jointly funded by Croud Inc Limited and Innovate UK under the Knowledge Transfer Partnership (KTP) Programme (Project Number: 10059992).

\section*{Conflict of Interests}
The authors declare that they have no conflicts of interest regarding the publication of this article. No financial or personal relationships that could have influenced the work have been disclosed.

\section*{Data Availability Statement}
Data not available due to commercial restrictions.

\section*{Appendix}

\section*{Appendix A: Constraits for underlying demand} 

The estimation of the underlying demand, $\boldsymbol U_t^\top \boldsymbol \beta_U$, within the model is subject to \textbf{explicit constraints} to ensure its commercial realism and statistical validity. This component, representing the baseline KPI in the absence of media influence, is not merely a fitted intercept. Instead, its magnitude is dynamically bounded within a plausible range, and its final estimated contribution is determined by its relative weighting within the overall model, particularly in relation to the incremental effects of media channels.

Beyond the overarching constraint on the total baseline's contribution, we impose specific constraints on the \textbf{coefficients of each variable group} comprising $\boldsymbol U_t$. This ensures that the influence attributed to temporal patterns, economic conditions, pricing, distribution, exogenous factors, and brand metrics remains within empirically and commercially reasonable bounds. For any sub-vector of the underlying demand variable $\boldsymbol{V}_t \subset \boldsymbol{U}_t$, and the corresponding coefficient sub-vector $\boldsymbol \beta_V$, we define a constrained range for its total contribution to the baseline:
\begin{equation}
\phi_V^{\text{min}}\le \boldsymbol V_t^\top \boldsymbol \beta_V \le \phi_V^{\text{max}},
\end{equation}
where $\phi_V^{\text{min}}$ and $\phi_V^{\text{max}}$ are the bounds of the constrained range. The factors that inform these constraints are derived from a qualitative and quantitative understanding of market dynamics, reflecting insights often gained from market research, business intelligence, and industry benchmarks. These factors guide the setting of these group-specific contribution ranges:

\begin{itemize}
    \item \textbf{Repeat Customer Rate}: A higher proportion of \textbf{repeat customers} indicates strong brand loyalty and product stickiness. This typically correlates with a larger and more stable baseline demand, suggesting that $\boldsymbol \beta_U$ components related to inherent market size or brand strength might be bounded higher. 
    \item \textbf{Brand Recognition and Equity}: Elevated levels of \textbf{brand recognition} and established brand equity suggest a stronger intrinsic market presence. This leads to a larger baseline, as consumers are more likely to consider or choose the brand organically. A robust value in brand equity would generally allow for higher maximum contributions from stable underlying demand components.
    \item \textbf{Industry Competition}: A more \textbf{competitive industry landscape} can dilute individual brand baseline sales, as consumers have more alternatives and less inherent loyalty without active marketing. This implies that increased competition  generally corresponds to a smaller baseline for any given brand within that market, potentially narrowing the upper bounds for several $\boldsymbol \beta_U$ components.
    \item \textbf{Seasonal nature of Sales}: Products or services with a pronounced \textbf{seasonal sales pattern}, which is captured by $\boldsymbol S_{\text{seasonality},t}$ in $\boldsymbol U_t$, may exhibit a larger baseline during peak seasons due to inherent consumer demand cycles. The strength of this inherent seasonal effect informs the range for coefficients within $\boldsymbol S_t$, ensuring they reflect realistic seasonal amplitudes.
    \item \textbf{Online vs. Offline Sales Share}: The relative \textbf{share of online versus offline sales} can influence baseline estimations due to differing consumer behaviors and market structures across channels. Specifically, a greater proportion of online sales, often characterized by more direct comparisons, greater price transparency, and lower switching costs, might lead to a smaller \textit{online sales baseline} or different permissible ranges for price and promotional elasticities. This distinction acknowledges the unique market dynamics of digital channels.
\end{itemize}

These qualitative factors guide the setting of the quantitative ranges for each component group's contribution to the underlying demand. This process prevents the GA from assigning an implausible proportion of sales to any specific non-media driver, thereby ensuring that the optimization is constrained within a realm of feasible and realistic market dynamics. The specific constraints used to satisfy each of the underlying demands in this study are commercially sensitive and proprietary. As such, the detailed formulations are not disclosed in this paper to protect confidential commercial information.

\section*{Appendix B: Constraits for media channel parameters}

Constraints on media parameters are crucial for ensuring that the estimated contributions and response shapes of various marketing channels are both statistically sound and commercially realistic. These constraints are informed by a multi-faceted approach, integrating external market data, internal effectiveness assumptions, and observed channel performance. This approach reflects the need to ground statistical models in real-world marketing knowledge. The constraints are systematically applied to various parameters associated with the media channels $m \in \{1, \ldots, M\}$:

\begin{enumerate}
    \item \textbf{Prioritization based on Media Consumption by Channel and Age Demographic:}
    To establish initial ranges for the relative contribution and diminishing returns of media channels, we first analyze media consumption patterns across various channels and age demographics. This demographic-level understanding provides a foundational insight into the potential reach and engagement capacity of each channel, reflecting principles of audience targeting in media planning. For instance, traditional media channels like linear TV and Radio exhibit higher consumption rates among older generations, suggesting their primary effectiveness lies within these demographic segments. Conversely, Paid Social media platforms are predominantly consumed by younger generations, indicating a stronger potential impact within these cohorts. Even within Paid Social, platforms demonstrate distinct demographic leanings (e.g., TikTok usage is heavily skewed towards younger audiences compared to Facebook), necessitating differential weighting or constraining based on target audience alignment. This qualitative assessment guides the relative bounds for channel-specific parameters.

    \item \textbf{Weighted Media Activity (Effective Exposure):}
    The next step involves refining these initial insights by calculating weighted media activity, which accounts for both the volume of media exposure and its assumed effectiveness. This methodology aims to convert raw media spend or impressions ($X_{m,t}^{\text{raw}}$) into a more accurate representation of \textbf{effective exposure}.
    \begin{itemize}
        \item \textit{Effectiveness Weighting:} A critical factor is the effectiveness of each media channel in reaching and engaging the target audience. This effectiveness is often represented by an empirically derived or qualitatively assessed coefficient, $w_m \in (0, 1]$, reflecting factors like ad viewability, clutter, and format impact. For example, Paid Social advertisements might be assigned a higher effectiveness weight ($w_{\text{Paid Social}} > w_{\text{Display}}$) due to generally superior viewability and engagement metrics, despite similar cost per impression.
        \item \textit{Weighted Media Consumption (WMC):} The \textbf{weighted media consumption} metric, denoted as $C_m$, integrates target demographic profiles with channel-specific media consumption rates. This adjusts the raw activity value to reflect its effective delivery to the most relevant audience segments. Specifically, $C_m$ can be formulated as:
        $$ C_m = w_m \times \text{AudienceMatch}_m $$
        where $\text{AudienceMatch}_m$ quantifies how much of the total target audience is reached by each segment on each media channel. This provides a refined measure of effective reach for parameter initialization and range setting.
    \end{itemize}

    \item \textbf{Parameter Range Definitions:}
    Utilizing the insights from media consumption analysis and weighted media activity, we establish precise minimum and maximum ranges for the key parameters of the media effectiveness model, namely the contribution coefficients ($\beta_{X,m}$), adstock rates ($\alpha_m$), shape parameters ($\gamma_m$), and inflection points ($\kappa_m$). These ranges are crucial for channeling the genetic algorithm's search towards economically and empirically plausible outcomes, acting as soft or hard bounds in the optimization process.

    \begin{itemize}
        \item \textbf{Media Contribution ($\beta_{X,m}$):} These define the upper and lower limits for the overall contribution of a given media channel to the KPI within the main model \eqref{eq:model}. These bounds, $[\beta_{X,m}^{\text{min}}, \beta_{X,m}^{\text{max}}]$, are informed by historical performance, industry benchmarks, and expert judgment, ensuring that a channel cannot be assigned an implausibly high or low incremental impact. The genetic algorithm searches within this defined interval.
        
        \item \textbf{Adstock Rate ($\alpha_m$):} For the adstock function defined in \eqref{eq:adstock}, the decay rate $\alpha_m$ is constrained to an industry-standard range, denoted by $\left[\alpha_m^{\text{min}}, \alpha_m^{\text{max}}\right] = [0.3, 0.6]$. This range reflects empirically observed decay patterns for advertising effects, preventing the model from fitting unrealistically short or excessively long carryover periods. 
        
        \item \textbf{Shape Parameter ($\gamma_m$):} For the Hill function defined in \eqref{eq:hill_function}, the shape parameter $\gamma_m$ is constrained to $\left[\gamma_m^{\text{min}}, \gamma_m^{\text{max}}\right] = [2.5, 3.5]$. This range captures typical diminishing-returns profiles, favoring S-shaped curves with a moderate “warming-up” phase followed by saturation, consistent with many advertising contexts. It also helps maintain numerical stability during optimization.

        \item \textbf{Inflection Parameter ($\kappa_m$):} The inflection parameter $\kappa_m$ in the Hill function \eqref{eq:hill_function} is critical for defining the point of diminishing returns. Its range, $[\kappa_m^{\text{min}}, \kappa_m^{\text{max}}]$, is determined dynamically for each channel $m$ based on its \textbf{Weighted Media Consumption ($C_m$)} and predefined bounds on effective reach and frequency. Specifically, the range is derived by multiplying $C_m$ by minimum and maximum weekly effective reach and frequency metrics. Let $\text{MinFreq}_m$, $\text{MaxFreq}_m$, $\text{MinReach}_m$ and $\text{MaxReach}_m$represent the desired minimum and maximum effective weekly frequencies and reach for channel $m$. Then:
        $$ \kappa_m^{\text{min}} = C_m \times \text{MinFreq}_m \times \text{MinReach}_m $$
        $$ \kappa_m^{\text{max}} = C_m \times \text{MaxFreq}_m \times \text{MaxReach}_m $$
        This ensures that the inflection point, representing the media activity level at which half of the maximum effectiveness is achieved, is anchored to the channel's actual scaled activity volume and realistic audience exposure thresholds. This approach helps to prevent the model from assigning an inflection point that is either unrealistically low (implying saturation at negligible spend) or excessively high (suggesting infinite returns).
    \end{itemize}
\end{enumerate}

\bibliographystyle{chicago}
\bibliography{References_MMM}

\end{document}